\documentclass[PAPER, cernpreprint, atlasdraft=false, USenglish, texmf, orcidlogo]{atlasdoc}
\PreprintIdNumber{CERN-EP-2024-154}
\usepackage{atlaspackage}
\usepackage{atlasbiblatex}

\usepackage{atlasphysics}
\usepackage{siunitx}

\graphicspath{{logos/}{figures/}}
\Crefname{figure}{Figure}{Figures}

\usepackage{ANA-TOPQ-2019-31-PAPER-defs}

\AtlasTitle{%
Search for heavy right-handed Majorana neutrinos in the decay of top quarks produced in proton--proton collisions at $\sqrt{s} = 13~\TeV$ with the ATLAS detector
}

\AtlasAbstract{%
A search for heavy right-handed Majorana neutrinos is performed with the ATLAS detector at the CERN Large Hadron Collider, using the 140~\ifb\ of proton--proton collision data at $\sqrt{s} = 13~\TeV$ collected during Run 2.
This search targets \ttbar production,
in which both top quarks decay into a bottom quark and a $W$ boson, where one of the $W$ bosons decays hadronically and the other decays into an electron or muon and a heavy neutral lepton.
The heavy neutral lepton is identified through a decay into an electron or muon and another $W$ boson, resulting in a pair of same-charge same-flavor leptons in the final state.
This paper presents the first search for heavy neutral leptons in the mass range of 15--75~\GeV using \ttbar events.
No significant excess is observed over the background expectation, and upper limits are placed on the signal cross-sections. Assuming a benchmark scenario of the phenomenological type-I seesaw model, these cross-section limits are then translated into upper limits on the mixing parameters of the heavy Majorana neutrino with Standard Model neutrinos.
}

\author{The ATLAS Collaboration}

\AtlasRefCode{TOPQ-2019-31}

\AtlasJournal{PRD}
\AtlasJournalRef{\PRD 110 (2024) 112004}
\AtlasDOI{10.1103/PhysRevD.110.112004}

\hypersetup{pdftitle={ATLAS document},pdfauthor={The ATLAS Collaboration}}

\begin{document}

\maketitle

\tableofcontents

\section{Introduction}
\label{sec:intro}
%

The observation of flavor oscillations in neutrinos produced in the sun, the earth atmosphere, reactors and accelerators implies that neutrinos have non-zero masses~\cite{Bellini_2014,Pontecorvo:1957cp,PhysRevD.17.2369,Mikheev:1986wj}.
Over the past decades, there were several important developments at the theoretical and experimental
frontiers addressing the question of neutrino-mass generation, the simplest among them being the type-I seesaw mechanism~\cite{PTP.64.1103} that defines three heavy neutral lepton (HNL) candidates, which are heavy right-handed Majorana neutrinos. These heavy neutrinos are their own anti-particles, and processes involving such a particle can violate lepton number ($L$) conservation with $|\Delta L|=2$.

Before the Large Hadron Collider (LHC)~\cite{Evans:2008zzb}, the best direct constraints in the HNL mass range 2--75~\GeV\ came from an analysis of the data collected by the DELPHI Collaboration at the LEP collider using $3.3 \times 10^6$ hadronic $Z$ boson decays~\cite{DELPHI:1996qcc}.
At the LHC, the most stringent limits on the HNL production cross-section come from searches for an HNL in \Wboson boson decays.
The CMS Collaboration explored the mass range 50~\GeV--25~\TeV\ using 138~\ifb\ of $\sqrt{s} = 13~\TeV$ proton--proton ($pp$) collision data in the same-charge dimuon final state, which provides the most stringent upper limits for the mass range above 650~\GeV~\cite{PhysRevLett.131.011803}.
The CMS Collaboration also searched for the decay products of long-lived HNLs as hadronic and electromagnetic showers detected in the muon chambers of the CMS detector~\cite{CMS-EXO-22-017}, a distinctive signature to probe for HNLs with masses lower than 4~\GeV\ and proper decay lengths of a few meters. This signature is sensitive to HNL couplings to all three generations of leptons, and the analysis set the most stringent limits to date on
the strength of HNL mixing with electron (muon) neutrinos
in the mass range of 2.1--3.0~(1.9--3.3)~\GeV.
In another analysis, the CMS Collaboration searched for both, Dirac and Majorana HNLs in the mass range of 10~\GeV--1.5~\TeV\ using three lepton final states~\cite{CMS-EXO-22-011}. This produced more stringent upper limits than the previously achieved experimental constraints for a wide range of HNL masses, and set limits on the strength of HNL mixing with $\tau$-neutrinos above the $W$ boson mass for the first time.

A previous result from the ATLAS Collaboration using 36.1~\ifb\ of $\sqrt{s} = 13~\TeV$ $pp$ collision data constrained the HNL mixing to muon and electron neutrinos in the mass range of 4.5--50~\GeV~\cite{ATLAS:jhep10.2019.265}. The ATLAS Collaboration recently searched for HNLs in $W$ boson decays using a dilepton displaced vertex, in the mass range of 3--15~\GeV~\cite{PhysRevLett.131.061803}. The latest ATLAS analyses searched for Majorana neutrinos in same-charge $WW$ scattering events through their coupling with electrons and muons in the HNL mass range 50~\GeV--20~\TeV~\cite{ATLAS:2023tkz,2403.15016}.

This is the first search for HNLs using $\ttbar$ events~\cite{ttbar_HNL}, where one top quark
decays into a bottom quark and a $W$ boson, with the $W$ boson further decaying into an electron or muon and a heavy neutral lepton.
The heavy neutral lepton decays promptly into another charged lepton and a $W$ boson, yielding a final state characterized by a pair of same-charge same-flavor leptons. The other top quark decays hadronically according to the Standard Model (SM), as illustrated in \cref{fig:feymanDsignal}. A detailed description of the signal model considered is provided in \cref{sec:signal}.

\begin{figure}[htb]
\centering
\includegraphics[width=0.5\textwidth]{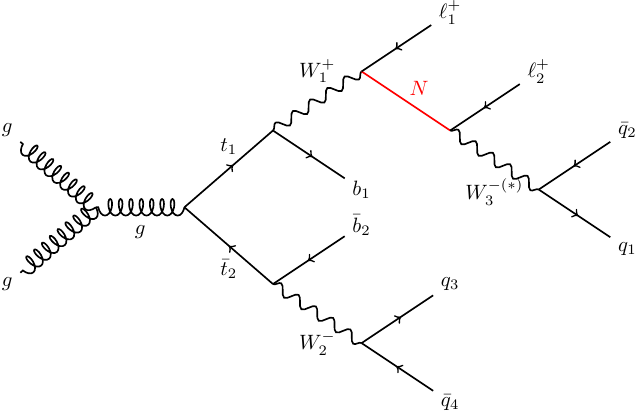}
\caption{
Example Feynman diagram of the signal process. The heavy neutral lepton is denoted by the symbol $N$. The charge-conjugate of this diagram is also a valid signal process. The off-shell \Wboson boson is marked with an asterisk.
}
\label{fig:feymanDsignal}
\end{figure}

Compared with the previous efforts in which HNLs are searched for in the decay of \Wboson bosons originating directly from the proton--proton collisions,
the advantage of using a \Wboson boson from a \ttbar event is the possibility of identifying one of the top quarks through its decay according to the SM, while utilizing the decay of the other top quark to search for HNLs.
Although the same signal model is expected to lead to fewer events in this channel, the background is also expected to be much smaller because of the state-of-the-art $b$-tagging algorithms. Because of the very different final states, this analysis provides complementary sensitivity to the previous searches.

Final states that have two same-charge electrons ($ee$ channel) or muons ($\mu\mu$ channel) are targeted.
A multivariate analysis is performed in combination with a statistical inference to extract the signal.
No significant excess is observed over the background expectation, and upper limits on the strength of the HNL mixing with electron and muon neutrinos are set in the mass range of $15~\GeV<m_\mathrm{HNL}<75~\GeV$, using 140~\ifb\ of $pp$ collision data collected by the ATLAS detector at $\sqrt{s}=13~\TeV$. The $ee$ and $\mu\mu$ channels are also used to constrain the strength of HNL mixing with $\tau$-neutrinos, considering purely leptonic decays of $\tau$-leptons. While calculating the upper limits on one mixing parameter, the other two are assumed to be zero.

The paper is structured as follows.
\cref{sec:detector} provides a brief overview of the ATLAS detector used to collect the data.
\cref{sec:datasets} describes the signal model, the background processes, the detector simulation, and the theoretical predictions considered.
The objects used for the analysis, the event selection, and the event reconstruction are described in \cref{sec:objects_and_selection}.
The signal and control regions are defined in \cref{sec:SR_CR}, which also describes the multivariate analysis used for the signal extraction.
Systematic uncertainties are described in \cref{sec:systematics}.
Finally, \cref{sec:results} describes the statistical method used to analyze the experimental data, and presents the final results, while \cref{sec:conclusions} presents the conclusion.


%
\section{ATLAS detector}
\label{sec:detector}
%

%
\newcommand{\AtlasCoordFootnote}{%
ATLAS uses a right-handed coordinate system with its origin at the nominal interaction point (IP)
in the center of the detector and the \(z\)-axis along the beam pipe.
The \(x\)-axis points from the IP to the center of the LHC ring,
and the \(y\)-axis points upwards.
Polar coordinates \((r,\phi)\) are used in the transverse plane,
\(\phi\) being the azimuthal angle around the \(z\)-axis.
The pseudorapidity is defined in terms of the polar angle \(\theta\) as \(\eta = -\ln \tan(\theta/2)\) and is equal to the rapidity
$ y = \frac{1}{2} \ln \left( \frac{E + p_z c}{E - p_z c} \right) $ in the relativistic limit.
Angular distance is measured in units of \(\Delta R \equiv \sqrt{(\Delta y)^{2} + (\Delta\phi)^{2}}\).}

The ATLAS detector~\cite{PERF-2007-01} at the LHC covers nearly the entire solid angle around the collision point.\footnote{\AtlasCoordFootnote}
It consists of an inner tracking detector surrounded by a thin superconducting solenoid, electromagnetic and hadronic calorimeters,
and a muon spectrometer incorporating three large superconducting air-core toroidal magnets.

The inner-detector system (ID) is immersed in a \qty{2}{\tesla} axial magnetic field
and provides charged-particle tracking in the range \(|\eta| < 2.5\).
The high-granularity silicon pixel detector covers the vertex region and typically provides four measurements per track,
the first hit generally being in the insertable B-layer (IBL) installed before Run~2~\cite{ATLAS-TDR-19,PIX-2018-001}.
It is followed by the SemiConductor Tracker (SCT), which usually provides eight measurements per track.
These silicon detectors are complemented by the transition radiation tracker (TRT),
which enables radially extended track reconstruction up to \(|\eta| = 2.0\).
The TRT also provides electron identification information
based on the fraction of hits (typically 30 in total) above a higher energy-deposit threshold corresponding to transition radiation.

The calorimeter system covers the pseudorapidity range \(|\eta| < 4.9\).
Within the region \(|\eta|< 3.2\), electromagnetic calorimetry is provided by barrel and
endcap high-granularity lead/liquid-argon (LAr) calorimeters,
with an additional thin LAr presampler covering \(|\eta| < 1.8\)
to correct for energy loss in material upstream of the calorimeters.
Hadronic calorimetry is provided by the steel/scintillator-tile calorimeter,
segmented into three-barrel structures within \(|\eta| = 1.7\), and two copper/LAr hadronic endcap calorimeters.
The solid angle coverage is completed with forward copper/LAr and tungsten/LAr calorimeter modules
optimized for electromagnetic and hadronic energy measurements respectively.

The muon spectrometer (MS) has separate trigger and
high-precision tracking chambers measuring the deflection of muons in a magnetic field generated by the superconducting air-core toroidal magnets.
The field integral of the toroids ranges between \num{2.0} and \qty{6.0}{\tesla\metre}
across most of the detector.
Three layers of precision chambers, each consisting of layers of monitored drift tubes, cover the region \(|\eta| < 2.7\),
complemented by cathode-strip chambers in the forward region, where the background is highest.
The muon trigger system covers the range \(|\eta| < 2.4\) with resistive-plate chambers in the barrel, and thin-gap chambers in the endcap regions.

The luminosity is measured mainly by the LUCID--2~\cite{LUCID2} detector that records Cherenkov light produced in the quartz windows of photomultipliers located close to the beam pipe.

Events are selected by the first-level trigger system implemented in custom hardware,
followed by selections made by algorithms implemented in software in the high-level trigger~\cite{TRIG-2016-01}.
The first-level trigger accepts events from the \qty{40}{\MHz} bunch crossings at a rate below \qty{100}{\kHz},
which the high-level trigger further reduces to record complete events to disk at about \qty{1}{\kHz}.

A software suite~\cite{SOFT-2022-02} is used in data simulation, in the reconstruction
and analysis of real and simulated data, in detector operations, and in the trigger and data acquisition
systems of the experiment.


%
\section{Data and simulated event samples}
\label{sec:datasets}
%

\label{sec:MCttbar}
\label{sec:MCbkg}

\newcommand{\hdampFootnote}{%
The \hdamp parameter controls the matching between the matrix elements in \powheg and parton shower in \Pythia. It effectively regulates the high-\pt\ radiation against which the \ttbar\ system recoils.}

The search uses LHC $pp$ collision data at $\sqrt{s} = 13~\TeV$ collected with the ATLAS detector. Data quality requirements are applied to ensure that all detector components are fully operational~\cite{DAPR-2018-01}. The resulting data sample corresponds to an integrated luminosity of 140~\ifb, obtained using the LUCID-2 detector \cite{LUCID2} for the primary luminosity measurements, complemented by measurements using the inner detector and calorimeters~\cite{DAPR-2021-01}.

Simulated Monte Carlo (MC) events are used to model the signal and the SM backgrounds. The effects of the detector are incorporated using a simulation of the ATLAS detector~\cite{SOFT-2010-01} obtained with the \GEANT program~\cite{Agostinelli:2002hh} or with a faster approach employing parameterized showers in the calorimeter.
The effects of the additional $pp$ collisions in the same or a nearby bunch crossing (pile up) are modeled by overlaying minimum bias events simulated using \PYTHIA[8.1]~\cite{Sjostrand:2007gs} with the A3 set of tuned parameters (tune) on events from hard-scatter processes~\cite{ATL-PHYS-PUB-2016-017}. The MC events are weighted to reproduce the distribution of the average number of interactions per bunch crossing observed in the data.
All samples simulated with \POWHEGBOX~\cite{Frixione:2007nw,Nason:2004rx,Frixione:2007vw,Alioli:2010xd}, and \MG~\cite{Alwall:2014hca} are interfaced to \Pythia[8] to simulate the parton shower, fragmentation, and underlying event with the A14 tune~\cite{ATL-PHYS-PUB-2014-021} and the \NNPDF[2.3lo]~\cite{Ball:2012cx} parton distribution function (PDF) set.
Samples using \Pythia[8] and \Herwig[7]~\cite{Bahr:2008pv,Bellm:2015jjp} have heavy-flavor hadron decays modeled by \EVTGEN~\cite{Lange:2001uf}. The masses of the top quark, $m_t$, and of the Higgs boson, $m_H$, are set to 172.5~\GeV\ and 125~\GeV, respectively.
The data and simulated events are processed with the same reconstruction and analysis algorithms.

\subsection{Signal modeling}

\label{sec:signal}
\cref{fig:feymanDsignal} shows an example Feynman diagram of the signal process, with the final state used in this search.
The signal is a beyond the Standard Model (BSM) process, and it is simulated using an extension to the Minimal Standard Model, called $\nu$MSM~\cite{ASAKA2005151}, which is implemented as a universal FeynRules output (UFO) model called \texttt{HeavyN}~\cite{Degrande_2016}.

A simplified benchmark model is assumed that proposes a single HNL candidate $N$, which couples with either electrons, muons, or $\tau$-leptons through a mixing with the corresponding neutrinos.
In this model, the mass $m_N$ and couplings $V_{\ell,N}$ of this new particle are free parameters.
These couplings can be expressed through a mixing matrix,

\begin{equation}
V_{\ell,N}=
\begin{pmatrix}
V_{e,N} & 0 & 0\\
0 & V_{\mu,N} & 0\\
0 & 0 & V_{\tau,N}\\
\end{pmatrix},
\end{equation}

where only one of the three diagonal terms can be non-zero at a time.

The width of the HNL is automatically computed by the \MG~\cite{Alwall:2014hca} generator,
which provides matrix elements at next-to-leading-order (NLO) in QCD
using the \NNPDF[3.0nlo]~\cite{Ball:2014uwa} PDF set, and uses the five-flavor scheme.
The QCD factorization ($\mu_\mathrm{F}$) and renormalization ($\mu_\mathrm{R}$) scales are set to be equal~\cite{Frederix2018}, given by the dynamical expression
$\frac{1}{4}\sum_{i}\sqrt{m^{2}_i+p^{2}_{\textrm{T},i}}$,
where the sum runs over all the particles generated in the matrix element calculation, and the mass and transverse momentum of each particle are denoted by $m_i$ and $p_{\textrm{T},i}$, respectively.
The decay of the top quarks is simulated
using the \madspin~\cite{Frixione:2007zp,Artoisenet:2012st} module, allowing three-body off-shell decays.
The parton shower, the hadronization, and the decay of the $\tau$-leptons are
simulated using \Pythia[8] with the A14 set of tuned parameters.

Signal samples are generated for ten values of $m_N$: 15, 25, 35, 40, 45, 50, 55, 60, 70 and 75 \GeV. The HNL decays promptly in this mass range. For each mass point, six separate samples are produced for the two possible choices in the sign of the same-charge leptons ($\ell^{+}\ell^{+}$ or $\ell^{-}\ell^{-}$) and three possible choices in their flavor ($\ell$ = $e$, $\mu$ or $\tau$-lepton).

%

%
%
%
%
%
%

%

%
%
%
%


\subsection{Background processes with prompt leptons}
\label{sec:backgrounds}

Leptons that originate from the primary hard-scattering event and satisfy all selection requirements, described in \cref{sec:event_selection}, are referred to as prompt leptons. Conversely, leptons that enter the final states by any background process other than $W$, $Z$ or Higgs boson decays are called non-prompt leptons. Samples of simulated events are produced to model the different background processes, along with additional samples to estimate the modeling uncertainties for each process.

The production of \ttbar\ events is modeled using the \POWHEGBOX[v2] generator that provides matrix elements at NLO precision with the \NNPDF[3.0nlo] PDF set and the \hdamp\ parameter\footnote{\hdampFootnote} set to 1.5~\mt~\cite{ATL-PHYS-PUB-2016-020}.
The functional form of the renormalization and factorization scale is set to the default scale $\sqrt{\mt^2 + p_{\textrm T}^2}$. The parton shower and hadronization are modeled with the \PYTHIA[8.2] generator.
The uncertainty in matching the NLO matrix elements to the parton shower when generating $t \bar{t}$ events is evaluated by comparing the nominal samples of simulated events to samples with an alternative setting of the $p_{\mathrm{T}}^{\text{hard}}$ parameter in the matching code, using $p_{\mathrm{T}}^{\text{hard}}=1$ instead of the default setting of $p_{\mathrm{T}}^{\text{hard}}=0$~\cite{höche2021study}. This parameter regulates the definition of the vetoed region of the parton shower and is important in avoiding overlap in the phase space filled by \POWHEG and \Pythia. The uncertainty in the choice of the $h_{\text{damp }}$ parameter for the $\ttbar$ event generation is estimated using an additional $t \bar{t}$ sample produced as the nominal sample, but with the $h_{\text{damp }}$ parameter set to $3~m_t$~\cite{top-quark}.
The uncertainty in the parton-shower and hadronization model for the \ttbar events is evaluated by comparing the nominal generator setup with a sample produced with the \POWHEGBOX[v2] generator and the \HERWIG[7.2.1] parton-shower and hadronization model. In this alternative sample, the \HERWIG[7] default set of tuned parameters~\cite{Bellm:2015jjp,Bellm:2017jjp} and the MMHT2014LO PDF set \cite{Harland-Lang:2014zoa} are used.
All \ttbar\ samples are normalized to the cross-section prediction at next-to-next-to-leading-order (NNLO) in QCD, including the resummation of next-to-next-to-leading logarithmic (NNLL) soft-gluon terms calculated using
the \textsc{Top++2.0}~program~\cite{Beneke:2011mq,Cacciari:2011hy,Baernreuther:2012ws,Czakon:2012zr,Czakon:2012pz,Czakon:2013goa,Czakon:2011xx}.
This cross-section corresponds to $\sigma(\ttbar)_{\textrm{NNLO+NNLL}} = \mathrm{832\pm51~fb}$.

The events for the \ttH process are simulated at NLO with \POWHEGBOX[v2] using the \NNPDF[3.0nlo] PDF set, while the Higgs-boson decay, hadronization, parton shower, and the underlying event are simulated with \PYTHIA[8.2]. The resulting sample is normalized to a cross-section of $507^{+35}_{-50}$~fb, which is calculated at NLO for both QCD and electroweak (EW) using \MG as reported in Ref.~\cite{deFlorian:2016spz}.
An additional \ttH sample is simulated with $p_{\mathrm{T}}^{\text{hard }} = 1$.
The uncertainty in the parton-shower and hadronization model is evaluated with the help of an alternative sample produced with \POWHEGBOX[v2] + \HERWIG[7.2.1], using the same prescription as the \ttbar process.

For the \ttW process, the \Sherpa[2.2.10]~\cite{Bothmann:2019yzt} generator and its default parton shower are used at NLO accuracy in QCD, with multi-leg merging of up to one additional parton at NLO and up to two additional partons at leading-order (\MEPSatNLO~\cite{Hoeche:2011fd,Hoeche:2012yf,Catani:2001cc,Hoeche:2009rj} set up with a merging scale of 30~\GeV).
Additionally, a leading-order (LO) EW sample also simulated with \Sherpa[2.2.10], but for the $\ttW+1~\mathrm{jet}$ final state, is used to simulate the EW corrections to the \ttW\ production.
Following the recommendations of the LHC Top and the LHC Higgs boson working groups, the cross-sections of these QCD and EW \ttW samples are scaled separately to $674.7~\mathrm{fb}$ and $47.7~\mathrm{fb}$, respectively~\cite{Frederix:2021agh}.
The impact of the systematic uncertainty in the choice of generator is evaluated using an alternative \ttW\ sample simulated with up to one additional parton in the final state at NLO accuracy in QCD using \MG. In this sample, the different jet multiplicities are merged using the FxFx NLO matrix-element and parton-shower merging prescription~\cite{Frederix:2021agh} with a merging scale of 30~\GeV. The events are interfaced with \PYTHIA[8.2].

The \ttZ production, with the \Zboson boson decaying leptonically, is simulated with \MG, interfaced with \PYTHIA[8.2]. Three samples corresponding to the three leptonic decay modes of the $Z$ boson are produced, and they are normalized to the total \ttZ production cross-section of $0.84^{+0.09}_{-0.10}$~pb~\cite{deFlorian:2016spz}.
The impact of the systematic uncertainty in the choice of generator is assessed through alternative \ttZ samples. The alternative sample addresses the uncertainty in the parton-shower and hadronization model by comparing the nominal generator setup with a sample produced using the \POWHEGBOX[v2] generator and the \HERWIG[7] parton-shower and hadronization model.

The $t \bar{t} \gamma$ process is simulated with \MG and the parton shower is simulated using \PYTHIA[8.2]. An overlap removal with the \ttbar sample is implemented based on the procedure used by the charge asymmetry analysis of the $t \bar{t} \gamma$ process~\cite{2023137848}. This sample corresponds to a total cross-section of 5.36~pb, as calculated by the MC generator at NLO.

The $tW$ production is modeled using the \POWHEGBOX[v2] generator at NLO in QCD using the five-flavor scheme and the \NNPDF[3.0nlo] set of PDFs.
The \PYTHIA[8.2] generator is used to model the parton shower and hadronization using the A14 tune and the \NNPDF[2.3lo] PDF set.

Single-top-quark $t$-channel production is modeled using the \POWHEGBOX[v2]generator at NLO in QCD using the four-flavor scheme and the corresponding \NNPDF[3.0nlo] PDF set.
Single-top-quark $s$-channel production is modeled using the \POWHEGBOX[v2]\ generator at NLO in QCD in the five-flavor scheme with the \NNPDF[3.0nlo]\ PDF set.
Both the samples are interfaced with \PYTHIA[8.2] to model parton-showering and hadronization.

In addition, the electroweak background ($V$ EW, where $V=W,Z$), composed of the $V$+jets, diboson ($VV$) and triboson ($VVV$) processes, is considered. The production of $V$+jets is simulated with the \SHERPA[2.2.1]
generator using NLO-accurate matrix elements for up to two jets, and LO-accurate matrix elements
for up to four jets calculated with the \textsc{Comix}~\cite{Gleisberg:2008fv}
and \textsc{OpenLoops}~\cite{Cascioli:2011va,Denner:2016kdg} libraries. They
are matched with the \Sherpa parton shower~\cite{Schumann:2007mg} using the \MEPSatNLO prescription
that uses the set of tuned parameters developed by the \Sherpa authors.
The \NNPDF[3.0nnlo]~\cite{Ball:2014uwa} PDF set is used, and the samples
are normalized to the NNLO prediction~\cite{Anastasiou:2003ds}.

Samples of diboson events are simulated with the
\SHERPA[2.2.1] or \SHERPA[2.2.2] generator, depending on the process.
Fully leptonic final states and semileptonic final states, where one boson
decays leptonically and the other hadronically, are simulated using
matrix elements at NLO accuracy in QCD for up to one additional parton
and at LO accuracy for up to three additional parton
emissions. Samples for the loop-induced processes $gg \to VV$ are
generated using LO-accurate matrix elements for up to one
additional parton emission for both the fully leptonic and
semileptonic final states. The matrix element calculations are matched
and merged with the \Sherpa parton shower based on Catani--Seymour
dipole factorization using the \MEPSatNLO prescription.
The virtual QCD corrections are provided by the \textsc{OpenLoops} library. The
\NNPDF[3.0nnlo] PDF set is used, along with a dedicated set of tuned parton-shower parameters developed by the \SHERPA authors.

The production of triboson events is simulated with the
\SHERPA[2.2.2] generator using factorized gauge boson decays.
Matrix elements, calculated at NLO for the inclusive process and at LO for up to
two additional parton emissions, are matched and merged with the \SHERPA parton
shower using the same prescription as the diboson sample. The virtual QCD correction for matrix elements is also performed similarly.

The $t\bar{t}ZZ$, $t\bar{t}WZ$, $t\bar{t}HH$, and $t\bar{t}WH$ processes are simulated using \MADGRAPH with the \NNPDF[2.3lo] PDF set, whereas the $t\bar{t}t$, $t\bar{t}t\bar{t}$, $tHW$\kern-1pt, $t\bar{t}WZ$, $t\bar{t}WW$\kern-1pt, $tHq$, $tWZ$, and $tZq$ samples are simulated using \MG with the \NNPDF[3.0nlo] PDF set. The $V H$ sample is simulated using \Pythia[8.2]. All these rare processes are normalized using their NLO theoretical cross-sections and are combined into the `Top other' category.

\subsection{Background processes with non-prompt leptons}
\label{sec:templates}

Besides prompt leptons, non-prompt leptons are also produced in many of the background processes, predominantly in \ttbar decays. Based on the information provided by the record of the simulated particles in the produced MC samples, each lepton is sorted into one of several lepton-origin categories. These categories distinguish between prompt leptons, electrons from prompt muon decay, electrons from photon conversion and leptons from the decay of a hadron. While selecting a simulated event, two same-charge same-flavor leptons ($ee$ or $\mu\mu$) are required in the final state, at least one of which is a prompt lepton. The following non-prompt categories are defined based on the origin of the second lepton.

The label `$t \bar{t} \ \gamma$-conv' refers to events from the $t\bar{t}\gamma$ and \ttbar samples in which the second lepton originates from photon conversion. Events from these two samples are merged after removing the $t\bar{t}\gamma$ events from the inclusive \ttbar sample.
The label `\ttbar Q-flip' refers to events from the \ttbar sample in which the second same-charge lepton has a misidentified charge, known as `charge-flip'.
The labels `\ttbar dec $e$' and `\ttbar dec $\mu$' indicate the background processes in which the second lepton originates from either a heavier lepton decay or a hadron decay.
Finally, if a simulated \ttbar event has a second lepton which does not correspond to any of the processes discussed above, it is included in the `Top other' category, mentioned in the previous section.


%
\section{Object reconstruction and event selection}
\label{sec:objects_and_selection}
%

The signal process involves \ttbar events, in which one of the top quarks decays hadronically according to the SM,
and the decay chain of the other top quark contains an HNL. The targeted final state consists of two same-charge light leptons (electrons or muons), two bottom quarks and four light quarks ($u$, $c$, $d$ or $s$-quark).

\subsection{Object reconstruction}
\label{sec:objects_reco}

The primary vertex of an event is defined as the three-dimensional point inside the beampipe that maximizes the $\sum{\pT^2}$ of the contributing tracks.

Electron candidates are reconstructed from energy deposits in the electromagnetic calorimeter matched to reconstructed tracks in the inner detector. The selected candidates are required to have transverse momenta \pT > 10~\GeV\
and pseudorapidity $|\eta| < 2.47$.
Candidates in the transition region between the barrel and endcap calorimeters, $1.37 \le \abseta \le 1.52$, have poorer energy resolution and are excluded.
To reduce the background from non-prompt sources, electron candidates are also required to satisfy $|d_0|/\sigma(d_0) < 5$ and $|z_0 \sin\theta| < 0.5~\mathrm{mm}$, where $d_0$ is the transverse impact parameter, with uncertainty $\sigma(d_0)$, and $z_0$ is the longitudinal impact parameter relative to the primary vertex.
The electron candidates are identified using a likelihood-based method and are required to satisfy the `TightLH' working point~\cite{EGAM-2018-01}. Leptons from heavy-flavor hadron decays, misidentified jets, or photon conversions are further
suppressed using a boosted decision tree (BDT) based discriminant~\cite{HIGG-2017-02}, which provides several isolation working points. The electrons are required to satisfy the loose isolation working point %
to ensure enough separation from nearby particles.
For some of the analysis regions discussed in \cref{sec:SR_CR}, a tighter isolation working point
is used for better rejection of non-prompt leptons.

The sign of the charge of an electron candidate is determined from the curvature of its track under the effect of the magnetic field present in the inner detector. For a highly energetic electron, this curvature can be very small, making it difficult to determine the direction in which the track bends, and consequently its electric charge.
Charge misidentification can occur if the electron emits a hard bremsstrahlung and subsequently produces an electron–positron pair through photon conversion, with one of these leptons having a high \pT. Typically, these conversions would be reconstructed as such, but in some asymmetric conversions only one of the tracks is reconstructed, and the charge of the corresponding electron candidate can be opposite of the charge of the original lepton that radiated the photon.
To deal with this challenge, a BDT-based tool~\cite{EGAM-2018-01} is employed, which combines several properties of an electron candidate, such as its charge, impact parameters, energy, and inner detector track, into a single discriminant.
For some of the analysis regions, the electrons are additionally required to satisfy this BDT based selection criterion for an improved rejection of electrons with a misidentified charge. This is referred to as the `electron charge identity selector' (ECIDS) tool.

A likelihood-based multivariate discriminant is constructed to enhance the selection of prompt electrons, while suppressing contributions from photon conversions and hadrons misidentified as electrons. It evaluates several properties of the electron candidates simultaneously, which include the shower shapes in the electromagnetic calorimeter, the quality of the track, and the detection of transition radiation in the TRT~\cite{EGAM-2018-01}. This is referred to as the `$e/\gamma$ ambiguity removal'.

The reconstruction of muon candidates combines charged particle tracks in the inner detector with those found in the muon spectrometer~\cite{MUON-2018-03}. These tracks must be consistent with originating from the primary vertex. Muon candidates must satisfy the `Medium' \cite{MUON-2018-03} identification and the loose isolation criteria~\cite{HIGG-2017-02}.
Similar to the electrons, a tighter isolation criterion is also employed in the definition of some of the analysis regions, as discussed in \cref{sec:SR_CR}.
In addition, the muons must have impact parameters satisfying $|d_0/\sigma(d_0)| < 3$ and $|z_0 \sin\theta| < \SI{0.5}{\mm}$. Muon candidates are required to have $\pt > 10$ \GeV\ and $\abseta < 2.5$.

One of the possible semileptonic decays of a top quark is into a $\tau$-lepton. Only leptonically decaying $\tau$-leptons are considered via their decays into electrons or muons, which are reconstructed according to the criteria described above. Hadronically decaying $\tau$-leptons are reconstructed as jets, and treated as light jets during the event selection.

Jets are reconstructed using a particle flow algorithm~\cite{PERF-2015-09} by combining information from both the inner detector and the calorimeters.
The \antikt algorithm~\cite{Cacciari:2008gp,Fastjet} with a radius parameter of \(R=0.4\) is used. Jets are required to have $\pt>20~\GeV$ and $|\eta|<4.5$ and are calibrated as described in Ref.~\cite{JETM-2018-05}.
To reduce the effect of pile up, the jet-vertex tagger (JVT)~\cite{PERF-2014-03}, which identifies jets originating from the primary vertex, is applied to the jets with $\pt < 60~\GeV$ and $|\eta|<2.4$, using the tight working point.
Additionally, `forward jets' with $2.5<|\eta|<4.5$ and $\pt<120~\GeV$ are required to satisfy a dedicated forward-JVT WP of fJVT < 0.4, in addition to a timing requirement~\cite{ATL-PHYS-PUB-2019-026}. The inclusion of forward jets enhances the analysis sensitivity by increasing the signal (and background) yields by about 20\%.

Jets originating from the hadronization of bottom quarks are identified ($b$-tagged) via the DL1r algorithm~\cite{r21_dl1r}.
This algorithm uses deep neural networks exploiting the distinct features of $B$-hadrons in terms of track impact parameters and displaced vertices reconstructed in the inner detector.
The inputs to the DL1r network include discriminating variables constructed by a recurrent neural network, which exploits the spatial and kinematic correlations between tracks originating from the same $B$-hadron.
A jet is considered $b$-tagged if it satisfies the operating point corresponding to an 85\% average efficiency for $b$-quark jets in simulated $t\bar{t}$ events.

The missing transverse momentum, whose magnitude is denoted in the
following by \met, is defined as the negative vector sum of the \pT\ of the reconstructed
and calibrated objects in the event~\cite{JETM-2020-03}. This sum also includes the momenta of the inner detector tracks with $\pt > \SI{0.5}{\GeV}$ that are matched to the primary vertex but are not matched to any other reconstructed objects.

An overlap removal procedure is applied to reduce any double-counting of objects that are independently reconstructed. An electron candidate is rejected if it shares a track with a muon candidate or if it has a track overlapping with another electron candidate. If there are jets within $\Delta R(j,e) = 0.2$\ of an electron candidate, the closest one fulfilling this criterion is removed. If there are other jets within $\Delta R(j,e) = 0.4$\ of the electron candidate, the electron candidate is removed.
Jets that are within $\Delta R(j,\mu) = 0.2$\ of a muon candidate or share a track with that muon candidate are removed if they contain fewer than three tracks. After that, any muon candidate that is within $\Delta R(j,\mu) = 0.4$\ of the remaining jets is removed.

\subsection{Event selection and reconstruction}
\label{sec:event_selection}

Each selected event is required to have exactly two light leptons ($e$ or $\mu$) of the same charge and flavor. The events are selected using single-lepton triggers~\cite{ATLAS:2019dpa,ATLAS:2020gty}, where the leading lepton is required to satisfy the trigger selection criteria. The leading lepton is also required to have $\pT > 27~\GeV$, where the trigger efficiency is 99\%.
Jets in the central region ($\abseta < 2.5$) are required to have $\pT > 20~\GeV$, whereas jets in the forward region are required to have $\pT > 35~\GeV$. The $b$-tagged jets are required to have $\pT > 25~\GeV$.
Events are required to have at least two $b$-tagged jets and at least four jets without a $b$-tag.
Finally, the lepton pair is required to have an invariant mass $m_{\ell\ell} > 12~\GeV$ to suppress the background from the Drell--Yan process.

As illustrated in \cref{fig:feymanDsignal}, the specific signal process\footnote{The charge conjugate is also considered and implied throughout the paper.} considered is: $t_{1} \to W_{1}^{+} b_{1}, W_{1}^{+} \to \ell_{1}^{+} N, N \to W_{3}^{-}\ell_{2}^{+}$ and $\bar{t}_{2} \to W_{2}^{-} \bar{b}_{2}, W_{2}^{-} \to q_{3} \bar{q}_{4}$. The off-shell $W_{3}^{-}$ decays hadronically. This provides four constraints on the invariant masses of the following reconstructed particles: $t_{1}$, $\bar{t}_2$, $W_{1}^{+}$ and $W_{2}^{-}$.
All possible combinations of the available objects are considered, and the decay chain is reconstructed using the one which minimizes

$$
\chi^2 = \frac{(m(t_1) - m_t)^2}{\sigma_t^2} +  \frac{(m(W_1^{+}) - m_W)^2}{\sigma_W^2} +  \frac{(m(\bar{t}_2) - m_t)^2}{\sigma_t^2} +  \frac{(m(W_2^{-})- m_W)^2}{\sigma_W^2},
$$

where $m_t=172.76~\GeV$ and $m_W=80.379~\GeV$ are the top and $W$ masses, respectively~\cite{ParticleDataGroup:2020ssz}. The parameters $\sigma_t=32~\GeV$ and $\sigma_W=27~\GeV$ are the widths of the reconstructed top quark and $W$ boson mass distributions, obtained by averaging over all the signal samples the individual results of a Gaussian distribution fit to these masses.

The invariant mass of the $W_1$ boson, decaying into the heavy neutral lepton, offers a good separation between signal and background.
For the signal process, the distribution of this variable peaks near the \Wboson boson mass, as shown in \cref{fig:bsmWmass}. For background processes passing the signal selection criteria, the invariant mass calculated from the two same-charge leptons and two light jets coming from different parents is expected to follow a broad non-resonant distribution.
This is used as one of the input variables for the multivariate analysis training, as discussed in \cref{sec:mva}.


%
\section{Signal extraction}
\label{sec:SR_CR}
%

Although all three flavors of HNLs are considered in this search, only the processes with $ee$ or $\mu\mu$ as the same-charge leptons in their final states are used for signal extraction. This implies that, while searching for the signal process with $\tau$-leptons, only the processes where both the $\tau$-leptons decay into same-flavor lighter leptons are considered.

Several analysis regions are defined: one signal region (SR) for each of the $ee$ and $\mu\mu$ channels, four control regions (CR) for the $ee$ channel and two CRs for the $\mu\mu$ channel.
While the SRs are enriched in signal events, each CR is enriched in events from a specific background process and have less than 1\% signal contamination.
A multivariate analysis is performed on the events belonging to each SR to achieve further separation between signal and background.

The selection criteria for the different signal and control regions are described in the following and summarized in \cref{tab:regions_def}.

\subsection{Signal regions}

The signal regions for both the $ee$ and $\mu\mu$ channels are defined by applying the tighter isolation criteria described in \cref{sec:objects_reco} on the same-charge leptons, and requiring $m_{\ell\ell} < 80~\GeV$ to minimize the $Z \to ee$ background. Additionally, in the $ee$ channel, both electrons are required to satisfy the criteria of the ECIDS tool and the $e/\gamma$ ambiguity removal, which are applied to suppress the charge flip and photon conversion contamination, respectively.

\begin{table}[!htbp]
\caption{Definitions of different signal and control regions. The control regions are enriched in events from the following processes: \ttW, heavy-flavor (HF) fake, photon-conversion (PC), and charge-flip (CF).
The `$Z$ veto' is defined as $m_{ee} \notin [m_Z-10~\GeV, m_Z+10~\GeV]$.}
\label{tab:regions_def}
\centering
\begin{tabular}{lllll}
\toprule
\multicolumn{5}{c}{Analysis regions for the $ee$ channel} \\
\midrule
Region  & Tight isolation & $e/\gamma$ ambiguity removal & ECIDS criteria  & $m_{ee}$ requirement \\
\midrule
SR & Both $e$ & Both $e$ & Both $e$  & $< 80~\GeV$ \\
$t\bar{t}W$ CR & Both $e$   & Both $e$ & Both $e$ &  $> 100~\GeV$ \\
HF CR & At most one $e$ & Both $e$ & Both $e$ &  $> 70~\GeV$\\
PC CR & -- & At most one $e$ &  Both $e$  & $> 75~\GeV$ and $Z$ veto \\
CF CR & -- & Both $e$ & At most one $e$ & $> 60~\GeV$ and $Z$ veto\\
\bottomrule
\end{tabular}
\begin{tabular}{lll}
\\
\toprule
\multicolumn{3}{c}{Analysis regions for the $\mu\mu$ channel} \\
\midrule
Region  & Isolation criteria & $m_{\mu\mu}$ requirement \\
\midrule
SR & Both $\mu$ & $ < 80~\GeV$ \\
$t\bar{t}W$ CR & Both $\mu$ &  $ > 80~\GeV$ \\
HF CR & At most one $\mu$ & $ > 75~\GeV$ \\
\bottomrule
\end{tabular}
\end{table}

\subsection{Control regions}

Various background processes, with similar final states as the signal processes, can mimic their signature and enter the SRs. In both the $ee$ and $\mu\mu$ channels, the \ttW process is an important background. In the $ee$ channel, the SM \ttbar process also produces the same-charge electron pair in the final state through heavy-flavor decay, photon-conversion, or charge-flip. In the $\mu\mu$ channel, the \ttbar process produces the same-charge muon pair mostly through heavy-flavor decay, as the occurrence of charge-flip in the $\mu\mu$ channel is found to be negligible. To understand the impact of the background contributions, dedicated CRs are defined.

To define a CR for the \ttW\ background, events with $m_{\ell\ell} > 100~\GeV$ (in the $ee$ channel) or $m_{\ell\ell} > 80~\GeV$ (in the $\mu\mu$ channel) are selected, while keeping all other criteria of the SR unchanged. The tighter selection in the $ee$ channel reduces the contribution from the $Z \to e^+e^-$ process, which can produce a same-charge electron pair in the final state through charge-flip. A single \ttW CR is defined by merging events from both channels.

`Fake leptons' are reconstructed lepton candidates that originate either from real leptons, produced in non-prompt decays or photon conversion, or from jets with an electromagnetic signature that mimics that of electrons. Fake leptons are often created when a hadron, containing a $b$- or $c$-quark, decays into a lighter hadron and a lepton. A CR for such heavy flavor fake leptons is defined by selecting events in which at least one of the leptons does not satisfy the tighter isolation criteria described in \cref{sec:objects_reco}, along with a requirement of $m_{\ell\ell}>70~\GeV$ (for the $ee$ channel) or $m_{\ell\ell}>75~\GeV$ (for the $\mu\mu$ channel) to minimize the signal contamination.

A CR with an enhanced contribution of the electron charge-flip background is defined by requiring at least one of the electrons to satisfy the reversed ECIDS criterion. In this CR, the contribution from the $Z \to ee$ process is mitigated by rejecting the events with $m_{\ell\ell}$ in a 10~\GeV\ window around the $Z$ boson mass ($Z$ veto). A requirement of $m_{\ell\ell}>60~\GeV$ is also imposed to reduce the signal contamination.

Finally, another CR is defined for the electrons originating from the interaction of photons with the detector material, known as photon conversion, by inverting the $e/\gamma$ ambiguity removal criterion to enhance the contribution from this targeted process. Additionally, in this CR, a lower bound of $m_{\ell\ell}>75~\GeV$ and the $Z$ veto are applied to reduce the signal contamination and the background events originating from the $Z \to ee$ process.

\FloatBarrier


%
\subsection{Multivariate analysis}
\label{sec:mva}
%

A multivariate analysis is performed in the SRs to achieve further separation between signal and background. The \texttt{GradientBoostingClassifier} model from the \texttt{scikit-learn}~\cite{scikit-learn} package is used to train boosted decision trees, using binary cross-entropy as the loss function.

Two separate BDTs are trained in different mass regions to maximize the separation power across the whole mass range: one for the low mass region using the four signal MC samples generated with $m_N$ between 15~\GeV\ and 40~\GeV, and another for the high mass region using the six samples with $m_N$ between 45~\GeV\ and 75~\GeV. The same BDTs are used for the $ee$ and $\mu\mu$ channels, and trained using a mixture of signal samples of the first two HNL generations, along with all the nominal background samples described in \cref{sec:backgrounds}. The mixture of signal samples used for the BDT training is prepared by aggregating equal numbers of events from the simulated samples belonging to the mass range corresponding to that BDT. The search for the third generation of HNL utilizes the same BDTs as the ones used in the searches for the first two generations.

Each simulated MC event has an associated event weight, which is the product of the generator-level weight and various scale factors applied for the calibration of different reconstructed objects present in that event. While training the model, absolute values of these event weights are used to preserve the convexity of the loss function. Different background components are normalized to the same luminosity according to their SM cross-sections while preparing the background sample. The signal and background samples are normalized to the same number of weighted events. A two-fold cross-validation was performed by splitting the whole dataset based on odd and even event numbers to check for overtraining. Although there was no sign of overtraining, the events with odd event number are nevertheless evaluated using the model trained on events with even event number, and vice versa.

\begin{figure}[htbp]
\centering
\subfloat[]{\includegraphics[width=0.45\textwidth]{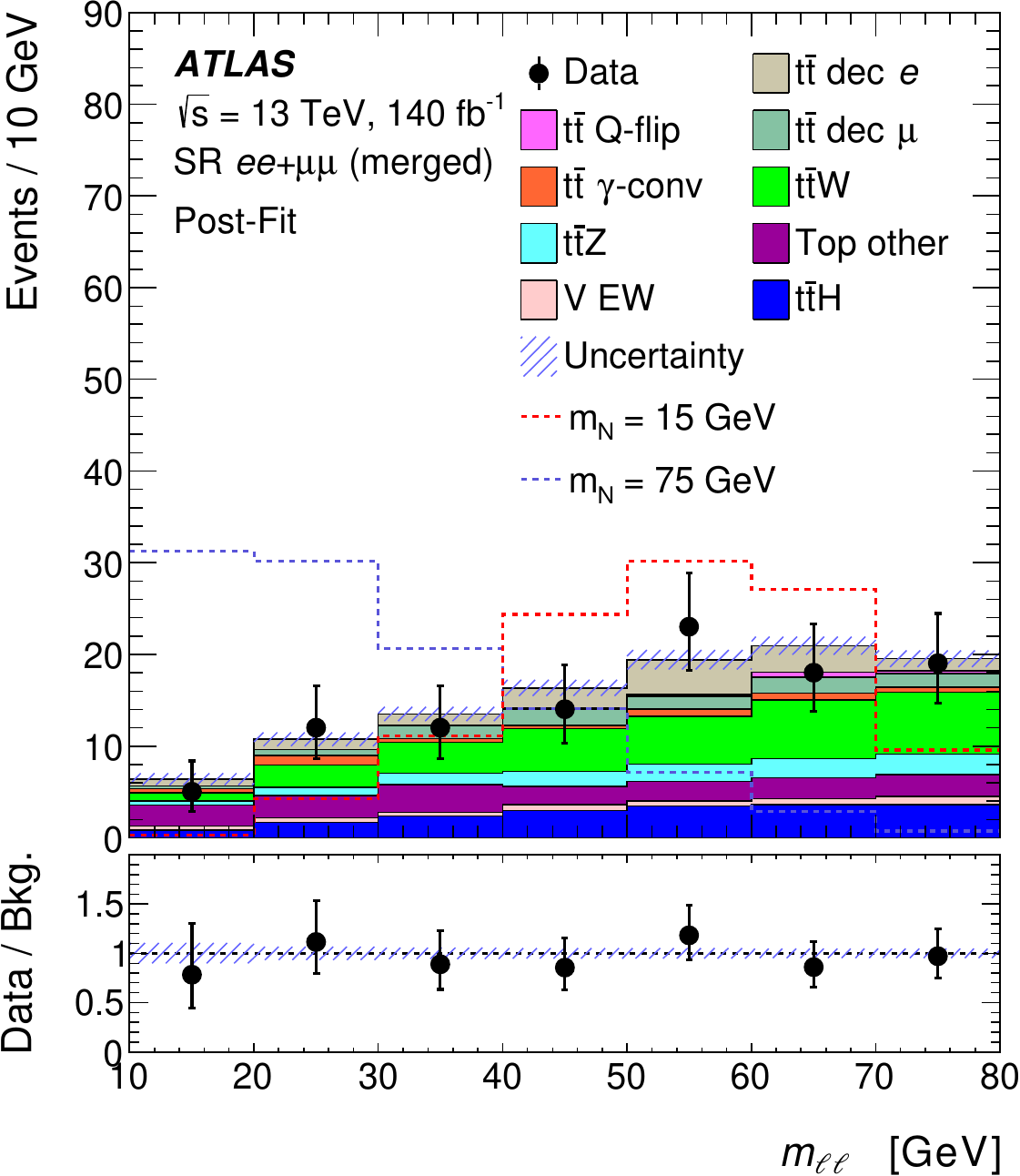} }
\hspace{0.05\textwidth}
\subfloat[]{\includegraphics[width=0.45\textwidth]{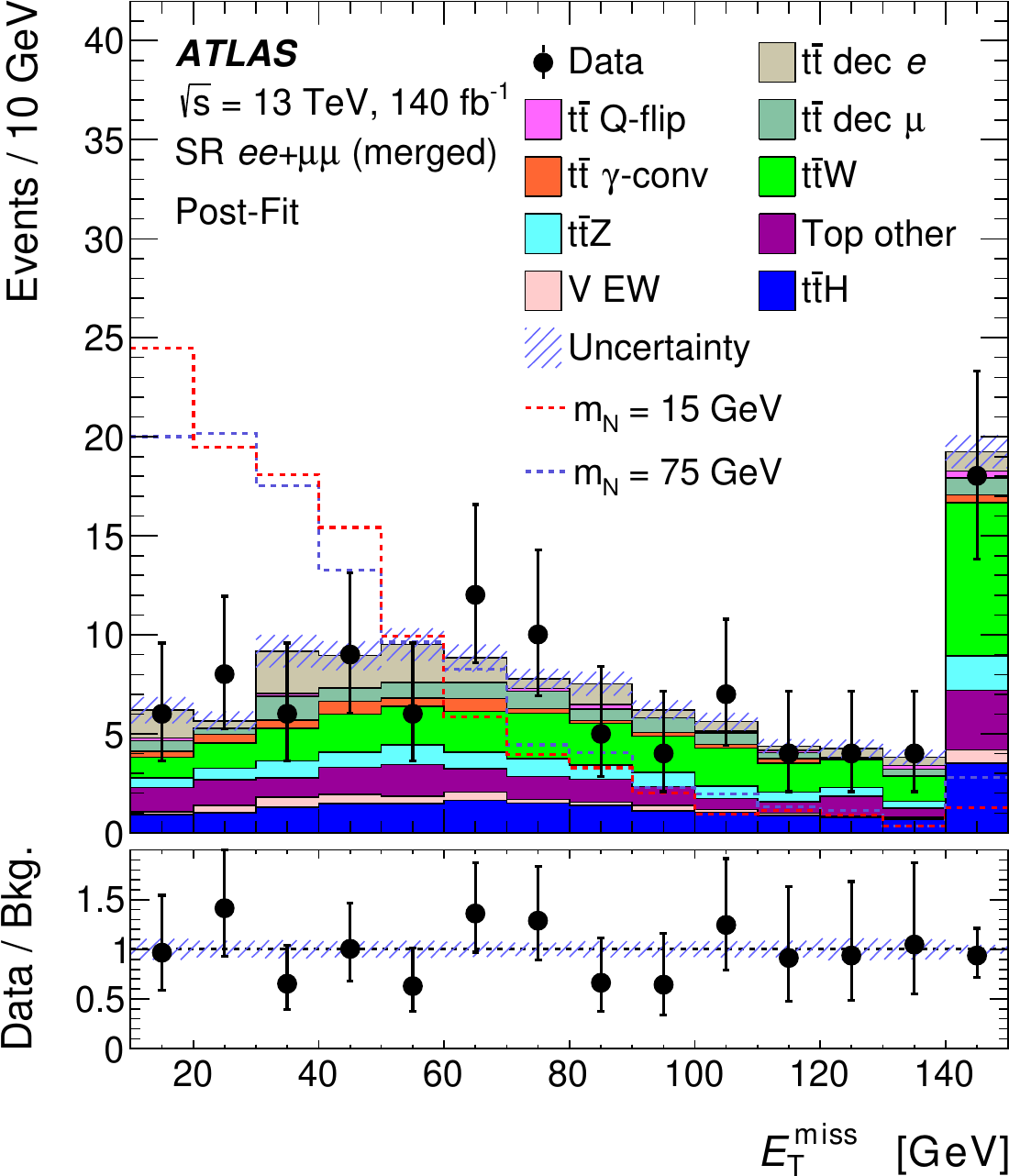} }
\\
\subfloat[]{\includegraphics[width=0.45\textwidth]{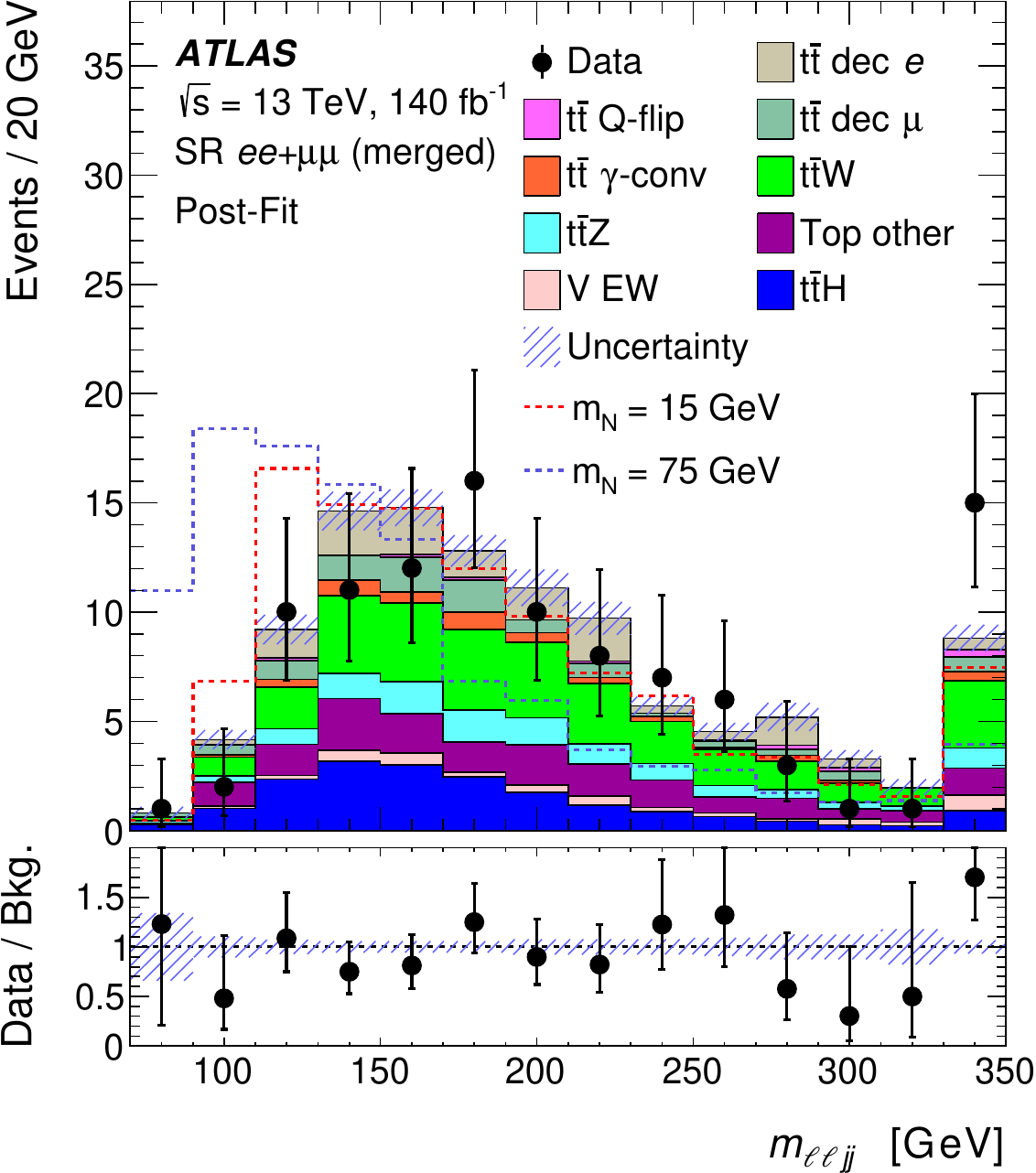}\label{fig:bsmWmass} }
\hspace{0.05\textwidth}
\subfloat[]{\includegraphics[width=0.45\textwidth]{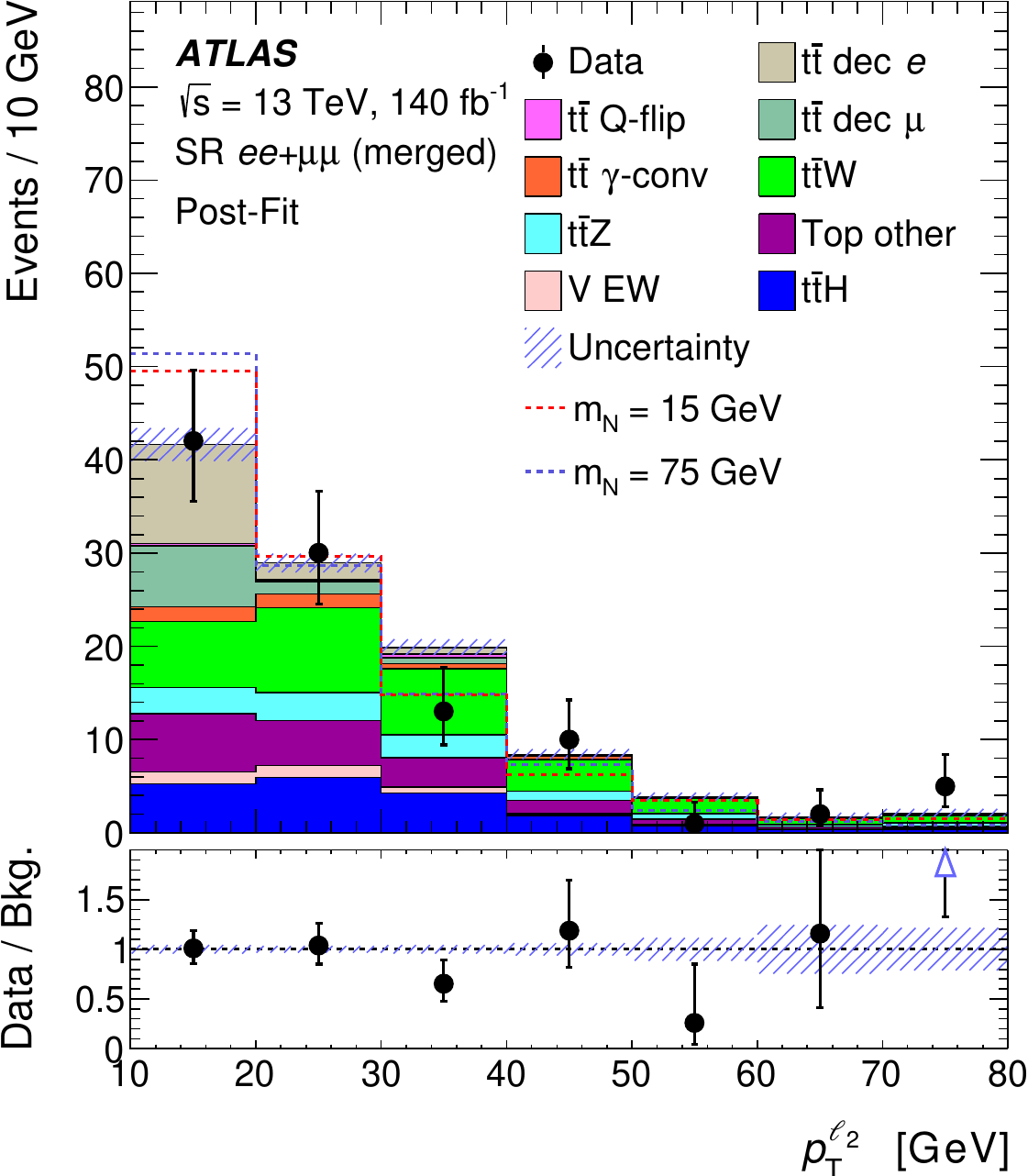} }
\caption{Distributions of the BDT input variables for two signal mass points and all background processes: (a)~invariant mass of the two same-charge leptons, (b) missing transverse momentum, (c) invariant mass of the $W_1$ candidate, and (d) transverse momentum of the sub-leading lepton. The normalizations of the background processes are obtained from a background-only fit to data across all the CRs in both channels. The signal distributions are normalized to the same number of events as the sum of all background processes.
The vertical error bars represent the statistical uncertainty in the data, shown as black dots, and the hatched band includes the MC statistical uncertainty and all the other systematic uncertainties added in quadrature. The first and last bins include the events below and above the histogram range, respectively.}
\label{fig:BDT_features}
\end{figure}

Based on optimization studies, the following input variables were found to be the most important (in decreasing order of importance), and are used to build the BDT models:
\begin{itemize}
\item Invariant mass of the two same-charge leptons ($m_{\ell\ell}$)
\item Missing transverse momentum of the event (\met)
\item Invariant mass of the \Wboson boson decaying into an HNL and a lepton ($m_{\ell\ell jj}$)
\item Transverse momentum of the sub-leading lepton ($\pt^{\ell_2}$)
\end{itemize}

The reconstruction of the second-last variable is discussed in \cref{sec:event_selection}. The distributions of the input variables for the two BDTs are shown in \cref{fig:BDT_features} for signal and background events separately. The signal distributions are normalized to the same number of events as the sum of all background processes to highlight the difference between their shapes. A satisfactory data-MC agreement is observed for all four input variables.

The separations obtained for two example mass points are illustrated in \cref{fig:N1_SR} (for the $ee$ channel) and \cref{fig:N2_SR} (for the $\mu\mu$ channel). The BDTs offer a better discrimination between signal and background than any of the input variables alone for all HNL mass points considered.


%
\section{Systematic uncertainties}
\label{sec:systematics}
%

The predictions of the signal and background processes are affected by theoretical uncertainties, and the data are affected by various sources of experimental uncertainties.
The uncertainties are classified into three different categories:
uncertainties arising from the reconstruction and calibration of the jet and lepton candidates, the uncertainty in the background
normalizations, and the uncertainty arising from the modeling of the signal and background processes.
They are detailed in the following subsections.

\subsection{Experimental uncertainties}

One of the primary sources of experimental uncertainty is the lepton efficiency, which is the efficiency of selecting an event containing the required number of leptons. This includes the reconstruction and identification efficiencies, the isolation efficiency, and the trigger efficiency. These efficiencies are different for data and simulated events, primarily because of the imperfect detector simulation. For simulated events, this is corrected by applying scale factors, calculated as the ratio of the efficiencies measured in the data to those measured in the simulation using the tag-and-probe method in $Z \to \ell\ell$ events. These scale factors, which depend on \pt and $\eta$, are varied up and down by one standard deviation to study the impact of the lepton efficiency uncertainties in the analysis result \cite{EGAM-2021-01,MUON-2018-03}. The uncertainty in the lepton trigger efficiency is calculated by propagating the uncertainties in individual trigger scale factors, which are derived in the same way as the identification and isolation scale factors for electrons and muons.

The lepton energy (momentum) measurements and correction factors are derived from a comprehensive study of dileptonic decays of the $Z$ boson, involving both the data and simulation. They are applied to correct possible mismodeling of the detector during the calibration. These correction factors are varied up and down by one standard deviation to study the lepton energy (momentum) scale uncertainty. The energy of the electrons is calibrated at the electromagnetic scale. For muons, the momentum scale and resolution corrections are applied to the MC simulations. Additional uncertainties are considered to account for the charge-dependent scale correction applied to data~\cite{PERF-2015-10}.

An uncertainty in the measurement of \met is assigned due to a possible miscalibration of its soft-track component, which are tracks that cannot be matched to any of the reconstructed and calibrated physics objects, i.e. hard components. They are derived from the level of agreement between data and simulation of the \pt balance between the hard and soft \met components~\cite{JETM-2020-03}. The scale and resolution uncertainties in the soft $\et^{\text{miss}}$ component are treated as separate contributions.

The overall uncertainty in the jet energy scale (JES) consists of individual components derived from in situ measurements, pile-up effects, flavor dependence, and additional effects like calorimeter responses and out-of-cone radiation. To determine the JES and its uncertainty, information from test-beam data, LHC collision data and simulation are used~\cite{JETM-2018-05}.

Precise knowledge of the jet energy resolution (JER) is important for the correct modeling of jet production. Thirteen orthogonal components account for jet $p_{\mathrm{T}}$- and $\eta$-dependent differences between simulation and data, which are determined using dijet events and noise measurements~\cite{JETM-2018-05}.

The uncertainties in JVT and fJVT are calculated by studying track-to-vertex matching~\cite{PERF-2014-03}. The $b$-tagging performance, characterized by the probability of correctly tagging a $b$-jet and the probability of mis-tagging a $c$-jet or a light flavor jet as a $b$-jet, are also taken into account~\cite{FTAG-2018-01,FTAG-2019-02,FTAG-2020-08}.

To account for the differences in the pile-up distribution between data and simulation, the pile-up profile in the simulation is corrected to match the one in data by applying several correction factors. The uncertainties in these correction factors are also taken into account.

Finally, the uncertainty in the measurement of the combined Run-2 integrated luminosity is estimated to be 0.83\%~\cite{DAPR-2021-01}.

\subsection{Background normalization uncertainties}

Uncertainties of 12\% and 10\% are included for the \ttZ and \ttH production cross-sections, respectively~\cite{deFlorian:2016spz}.
The \textit{$V$ EW} background consists of $Z$+jets events, with a 20\% uncertainty~\cite{Z+jets}, and $VV/VVV$ events, with a 50\%-60\% uncertainty~\cite{STDM-2018-03} in their cross-sections. Hence, an overall 30\% normalization uncertainty is considered for the \textit{$V$ EW} background as a weighted average.
For the $tW$, single-top-quark ($t$-channel and $s$-channel), and the minor backgrounds included in the `Top other' category, a 50\% normalization uncertainty is assigned.
The normalizations of the \ttW events, and of the \ttbar events with non-prompt leptons in their final states, are kept unconstrained and are determined by the fit, which is described in \cref{sec:results}.

\subsection{Modeling uncertainties}

The systematic uncertainties in MC modeling are estimated by comparing different generators or by varying the parameters used for the event simulation.

For the HNL signal samples, the following modeling systematic uncertainties are considered.
The uncertainty in the parton distribution functions is propagated using the PDF4LHC15 prescription~\cite{Butterworth:2015oua}.
The uncertainty due to missing higher-order corrections in the matrix element calculation is estimated by independently varying $\mu_\mathrm{R}$ and $\mu_\mathrm{F}$ up and down by a factor of two.

The dominant background is from the \ttbar events, in which at least one of the top quarks decays leptonically.
To estimate the uncertainty related to the choice of parton shower and hadronization model, the \powheg + \PYTHIA[8] sample is compared with the \powheg + \HERWIG[7] sample.
Independent variations of the renormalization and factorization scales are considered to account for missing higher-order corrections in the matrix element calculation.
The uncertainty in $\ttbar$ modeling due to the initial-state-radiation is estimated by varying the renormalization and factorization scales by factors of two around their original values, whereas the uncertainty in the modeling of final-state-radiation is evaluated by varying the renormalization scale for QCD emissions in the final-state shower by a factor of two around its original value.
The uncertainties in the cross-section due to the choice of PDF and the value of the strong coupling \alphas are calculated using the PDF4LHC prescription~\cite{Butterworth:2015oua} with the \MSTW[] 68\% CL NNLO~\cite{Martin:2009iq,Martin:2009bu}, CT10 NNLO~\cite{Lai:2010vv,Gao:2013xoa} and \NNPDF[2.3lo] PDF sets.
The uncertainty in matching the NLO matrix elements to the parton shower when simulating \ttbar events is evaluated by comparing nominal samples of simulated events with samples generated with an alternative setting of the $p_{\mathrm{T}}^{\text{hard}}$ parameter in the matching code, as discussed in \cref{sec:backgrounds}.
The uncertainty related to the choice of the $h_{\mathrm{damp }}$ parameter for the $\ttbar$ event simulation is estimated by using an additional $\ttbar$ sample produced with this parameter set to $3~m_t$, while keeping all other generator settings unchanged.

The nominal sample for the \ttH background is produced with the same generator as for the \ttbar background. Therefore, the uncertainties related to the parton shower, hadronization, renormalization and factorization scales, and the choice of $p_{\mathrm{T}}^{\text{hard }}$ parameter are estimated following the \ttbar recipe, except for the $h_{\mathrm{damp}}$ parameter, which is set to $3/4(m_t + m_{\bar{t}} + m_H) = 352.5~\GeV$, and is not varied.

For the \ttW\ background, the uncertainty related to the choice of parton shower and hadronization model is estimated by comparing the \powheg + \PYTHIA[8] sample with the \powheg + \HERWIG[7] sample.
To estimate the uncertainty related to the choice of generator, the nominal \SHERPA[2.2.10] sample is compared with the sample generated using \MG + \PYTHIA[8] with the FxFx merging scheme.
The uncertainties in scale variations are estimated from the independent variations of the renormalization and factorization scales by factors of two. The uncertainty in the parton distribution functions is propagated using the PDF4LHC15 prescription.

To estimate the uncertainty in the \ttZ background related to the choice of parton shower and hadronization model, the nominal sample is compared with the sample simulated using \MG + \HERWIG[7].
The uncertainties in scale variations are estimated from the independent variations of the renormalization and factorization scales by factors of two.
The uncertainties arising from initial-state-radiation and missing higher order terms in matrix element calculation are also accounted for.
%


%
\section{Statistical analysis and results}
\label{sec:results}
%

The signal strength $\mu$ is obtained from a simultaneous binned profile-likelihood fit~\cite{Cowan:2010js} across the signal and control regions, with systematic uncertainties included as nuisance parameters. The histograms used for constructing the likelihood with \texttt{HistFactory}~\cite{Cranmer:1456844} are the distributions of BDT output in the SRs, and are single-bin histograms with the event yields in the CRs. During the simultaneous fit, these single-bin histograms help to determine the normalizations of the particular background processes corresponding to the CRs.

The likelihood is constructed as a product of individual Poisson likelihoods over all considered bins.
The fitted event yields in the bins depend on nuisance parameters \(\theta\), which include the effects of systematic uncertainties.
Each nuisance parameter, except those representing the MC statistical
uncertainties, is constrained by a Gaussian distribution term in the likelihood function.
Some systematically varied discriminant distributions are smoothed and nuisance
parameters of systematic uncertainties with negligible impact are removed entirely to reduce spurious effects in minimization, improve convergence
of the fit, and reduce the computing time.
The effects of normalization and shape of a source of systematic uncertainty are
treated separately in this removal process.

The test statistic \(q_\mu\) is defined as the profile likelihood ratio
\[
q_\mu=-2 \ln \frac{
\mathcal{L}(\mu, \hat{\hat{\theta}})
}{
\mathcal{L}(\hat\mu, \hat\theta)
},
\]
where \(\hat\mu\) and \(\hat\theta\) are the values of the parameters that
maximize the likelihood function and $\hat{\hat{\theta}}$ are the values of the
nuisance parameters that maximize the likelihood function for a given value of
\(\mu\).
The test statistic is evaluated with the \texttt{RooStats} package~\cite{moneta2011roostats}
and is used in the modified frequentist ($\mathrm{CL}_S$)
method~\cite{A_L_Read_2002} to obtain upper limits on the signal strength.
For a given signal mass, the range of \(\mu\) yielding $\mathrm{CL}_S<0.05$, where
$\mathrm{CL}_S$ is computed using the asymptotic approximation~\cite{Cowan:2010js}, is
excluded at 95\% confidence level (CL).

While searching for an HNL mixing with a particular generation of SM neutrinos, the mixing with the other two generations is assumed to be zero. This implies, for an HNL mixing with the first two generations, the same-charge leptons in the final states of the signal processes directly couple to the HNLs and do not originate from $\tau$-lepton decays. Simultaneous fits across the signal and control regions are performed separately for the $ee$ and $\mu\mu$ channels to extract the signal strengths $\sigma_{e,N}$ and $\sigma_{\mu,N}$, respectively.
In the search for an HNL mixing with electron neutrinos, one SR and four CRs are included in the simultaneous fit.
The CRs used in the $ee$ channel are enriched in events from these background processes: $\ttW$\kern-1pt, fake electrons from heavy flavor hadron decays, charge flip and photon conversion.
In the search for an HNL mixing with muon neutrinos, one SR and two CRs are included in the simultaneous fit,
where the CRs are enriched in either $t\bar{t}W$ or fake muons from heavy flavor hadron decays.

The background-only fit distributions in the high mass region for the $ee$ and $\mu\mu$ channels are shown in \cref{fig:N1_SR_CRs} and \cref{fig:N2_SR_CRs}, respectively. The normalizations of the background components in the $ee$ channel, obtained from this fit, are $0.55\pm0.42$ for photon conversion, $1.52\pm0.69$ for heavy-flavor fake, $0.80\pm0.39$ for charge-flip, and $1.09\pm0.39$ for \ttW. In the $\mu\mu$ channel, the background normalizations are $0.75\pm0.32$ for heavy-flavor fake, and $1.05\pm0.41$ for \ttW. The net acceptance times efficiency for the signal processes in their corresponding SRs ranges from 0.26\% to 1.3\% depending on the HNL mass.

Similarly, while searching for an HNL mixing with the third generation, the same-charge electrons or muons in the final state of the signal process come from the $\tau$-lepton decays only. In this case, a simultaneous fit is performed across both the $ee$ and $\mu\mu$ channels.
The distributions from the background-only fit in the CRs are very similar to the ones obtained for the other two searches. Because of the lower number of expected signal events, a coarser binning has been used in the SRs of the $ee$ and $\mu\mu$ channels while searching for HNL mixing with $\tau$-leptons, as shown in \cref{fig:N3_SRs}.

\begin{figure}[htbp]
\centering
\subfloat[]{\includegraphics[width=0.45\textwidth]{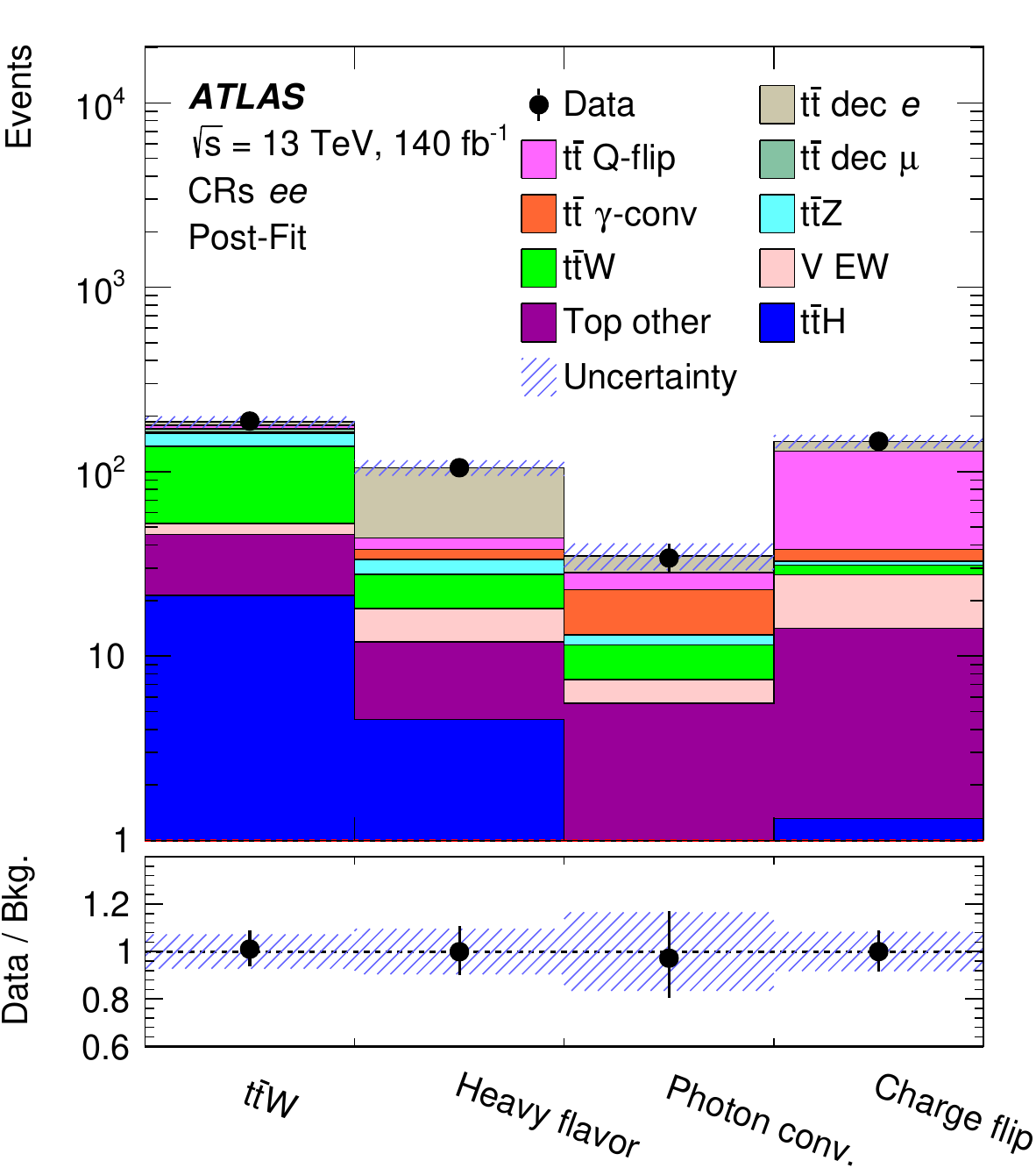}}
\hspace{0.05\textwidth}
\subfloat[]{\includegraphics[width=0.45\textwidth]{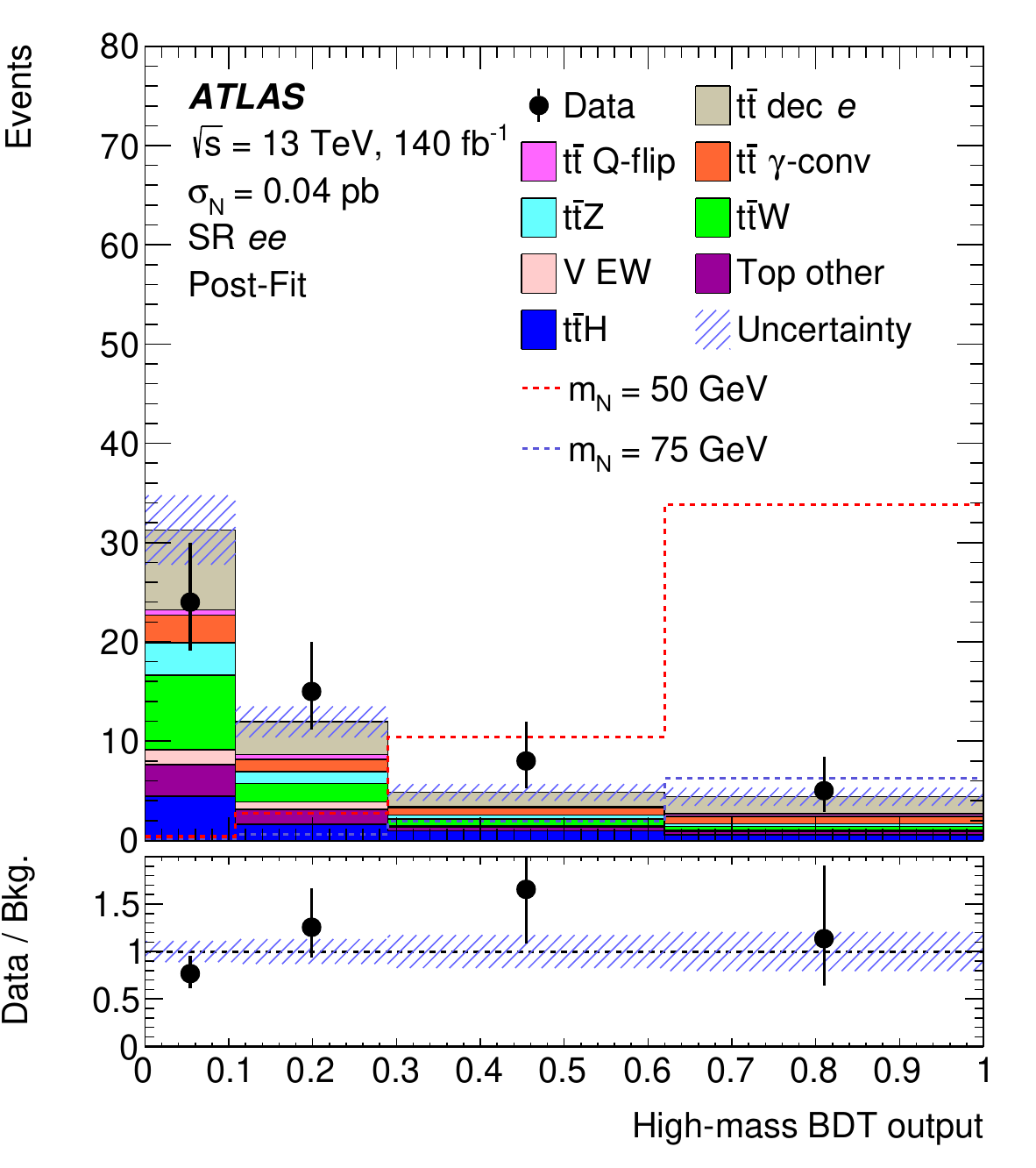}\label{fig:N1_SR}}
\caption{Data (dots) and MC (histograms) distributions, as obtained from the background-only fit in the $ee$ channel for
(a) the event yields from different background processes in the respective control regions, and (b) the distributions of the high-mass BDT output in the signal region for two signal mass points and all background processes.
The distributions for the signal processes correspond to the HNL mixing with electron neutrinos, and are overlaid on the stacked background distributions. Each of the signal distributions is normalized to a cross-section of 0.04~pb.
The vertical error bars represent the statistical uncertainty in the data, and the hatched band includes the MC statistical uncertainty and all the other systematic uncertainties added in quadrature.}
\label{fig:N1_SR_CRs}
\end{figure}

\begin{figure}[htbp]
\centering
\subfloat[]{\includegraphics[width=0.45\textwidth]{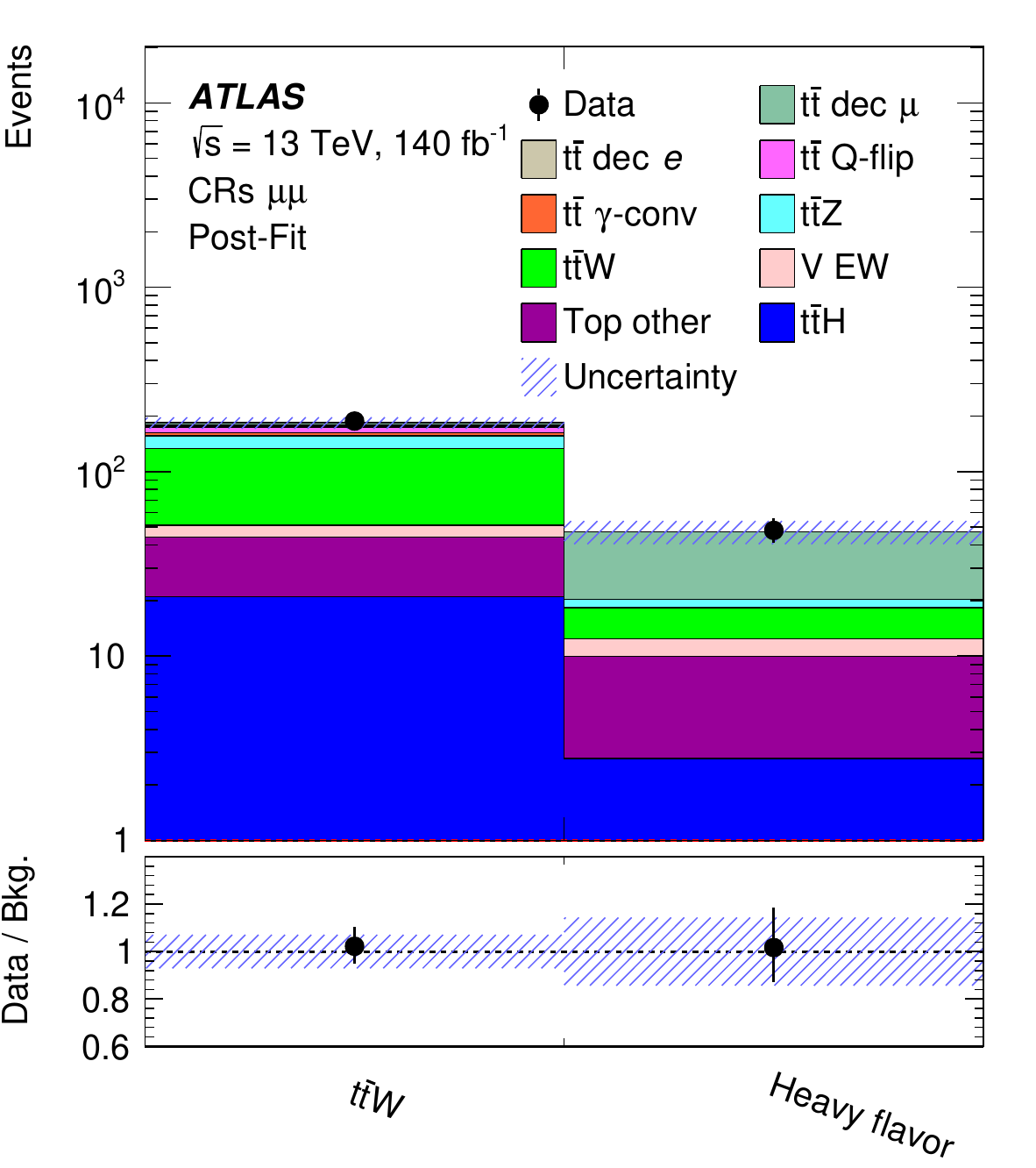}}
\hspace{0.05\textwidth}
\subfloat[]{\includegraphics[width=0.45\textwidth]{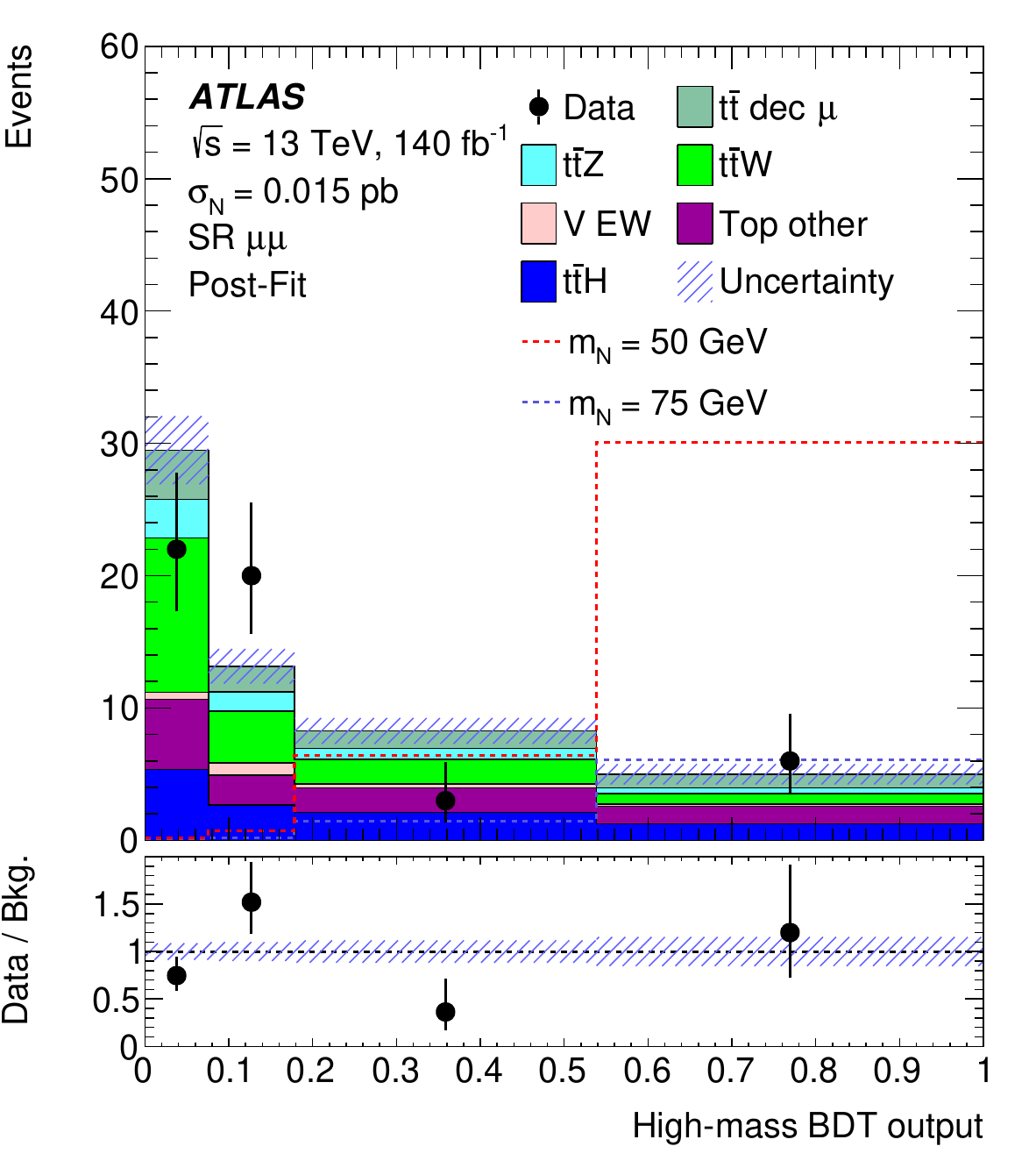}\label{fig:N2_SR}}
\caption{Data (dots) and MC (histograms) distributions, as obtained from the background-only fit in the $\mu\mu$ channel for
(a) the event yields from different background processes in the respective control regions, and (b) the distributions of the high-mass BDT output in the signal region for two signal mass points and all background processes.
The distributions for the signal processes correspond to the HNL mixing with muon neutrinos, and are overlaid on the stacked background distributions. Each of the signal distributions is normalized to a cross-section of 0.015~pb.
The vertical error bars represent the statistical uncertainty in the data, and the hatched band includes the MC statistical uncertainty and all the other systematic uncertainties added in quadrature.}
\label{fig:N2_SR_CRs}
\end{figure}

\begin{figure}[htbp]
\centering
\subfloat[]{\includegraphics[width=0.45\textwidth]{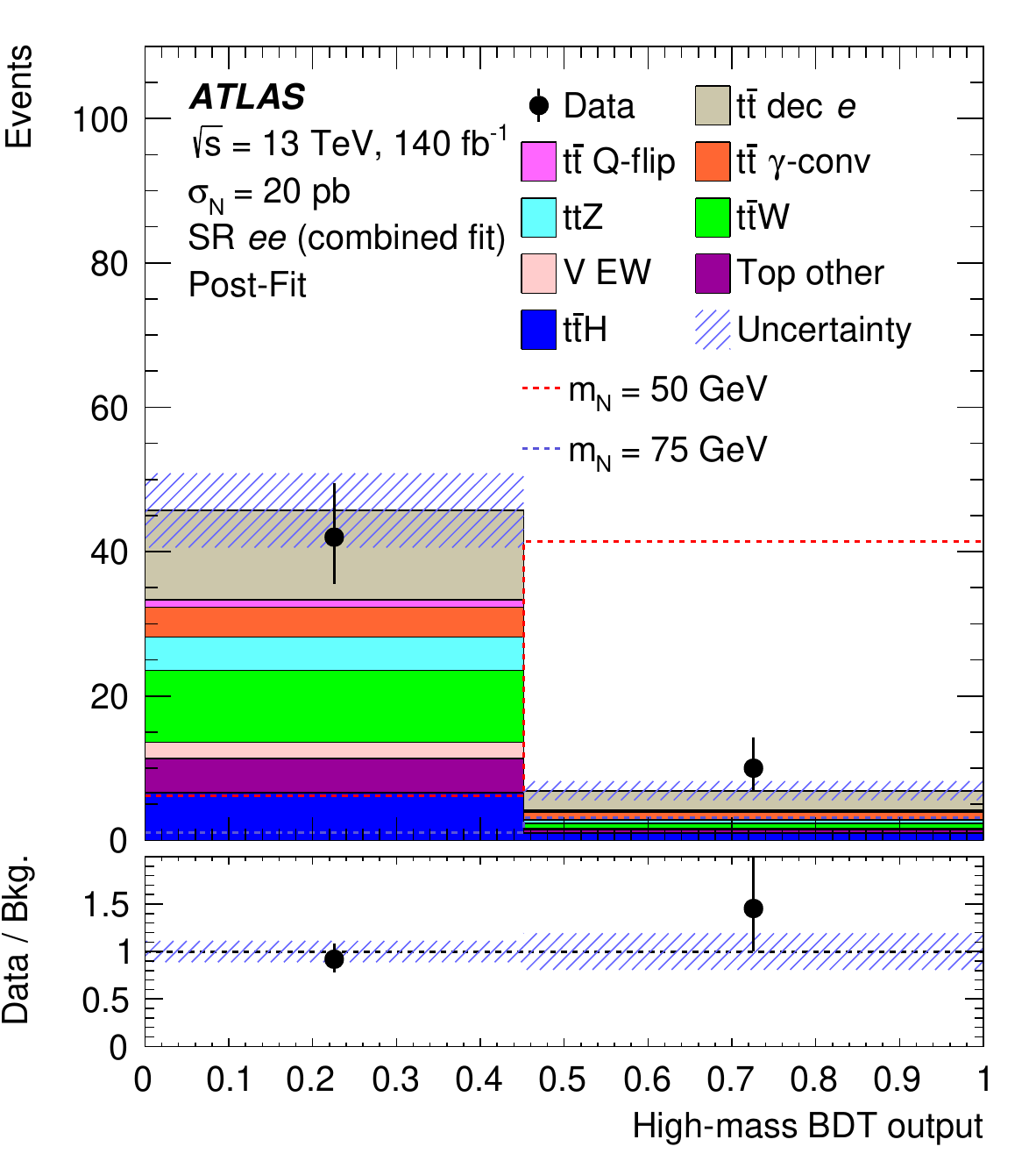}}
\hspace{0.05\textwidth}
\subfloat[]{\includegraphics[width=0.45\textwidth]{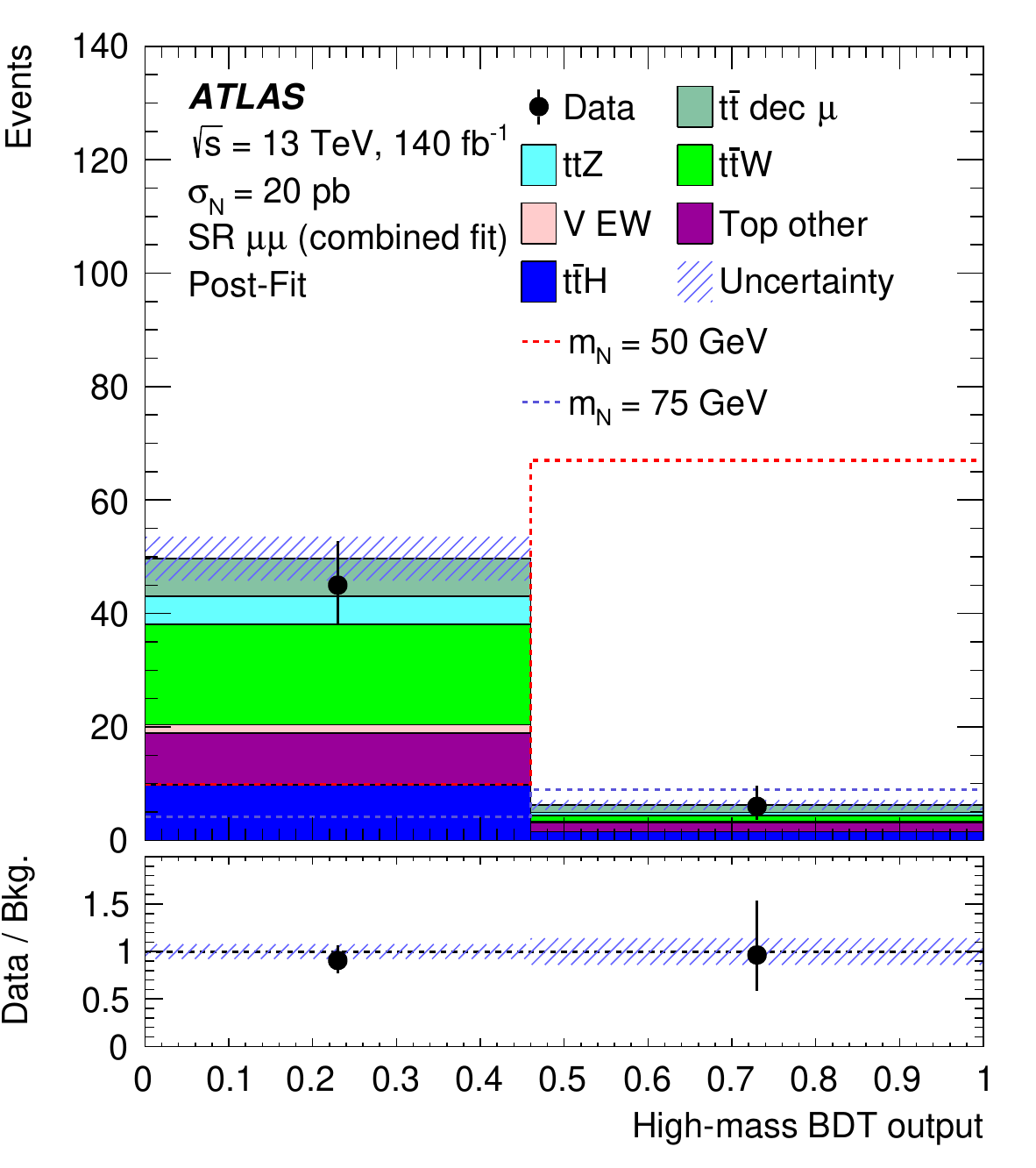}}
\caption{Data (dots) and MC (histograms) distributions of the high-mass BDT output, as obtained from the combined background-only fit, in the signal regions of the (a) $ee$ and (b) $\mu\mu$ channels, used to search for HNL mixing with $\tau$-neutrinos.
Distributions are shown for two signal mass points and all background processes.
The distributions for the signal processes correspond to the HNL mixing with $\tau$-neutrinos, and are overlaid on the stacked background distributions. Each of the signal distributions is normalized to a cross-section of 20~pb.
The vertical error bars represent the statistical uncertainty in the data, and the hatched band includes the MC statistical uncertainty and all the other systematic uncertainties added in quadrature.}
\label{fig:N3_SRs}
\end{figure}

\subsection{Upper limit on the signal strength}
The data and SM predictions agree within uncertainties, and no significant deviation is observed, as illustrated in \cref{tab:background_yields_unblind}. In the absence of any observed signal, upper limits are set on the signal strength at 95\% confidence level for different masses.
The upper limits on the cross-sections of the signal process are listed in \cref{tab:limit_unblind}. The most stringent limits are observed for $m_\mathrm{HNL}$ = 50~\GeV, for all three flavors of HNL.

The dominant sources of systematic uncertainties are the ones related to the modeling of the \ttbar process, normalization of the \ttW sample, and the JES/JER uncertainties. However, the effect of the systematic uncertainties on the expected and observed upper limits is small, as compared with its statistical counterpart. For $m_N=50~\GeV$, the expected upper limits on $\sigma_{e,N}$ and $\sigma_{\mu,N}$ change from 6.7~\ifb\ to 6.1~\ifb\ and from 3.1~\ifb\ to 2.8~\ifb, respectively, if the systematic uncertainties are not accounted for during the fit. This change is less than 10\% compared with the values obtained from the fits with all the systematic uncertainties included, for all the masses considered. The expected upper limit on $\sigma_{\tau,N}$ for the same mass decreases by 6\%.

\begin{table}[htbp]
\newcommand{\tabledash}{\multicolumn{2}{c}{--}}
\centering
\caption{Post-fit event yields for the different background processes in the signal regions, as obtained from the background-only fit in the high-mass region.}
\begin{tabular}{l *4{S[table-format=3.1]@{$\,\pm\,$}S}}
\toprule
& \multicolumn{4}{c}{Separate fits} & \multicolumn{4}{c}{Combined fit} \\
\midrule
& \multicolumn{2}{c}{SR $ee$} & \multicolumn{2}{c}{SR $\mu\mu$}
& \multicolumn{2}{c}{SR $ee$} & \multicolumn{2}{c}{SR $\mu\mu$} \\
\midrule
\ttbar decay $e$ & 15 & 6 & \tabledash & 15 & 7 & \tabledash \\
\ttbar Q-flip & 1.4 & 0.9 & \tabledash & 1.3 & 0.8 & \tabledash \\
\ttbar $\gamma$-conv & 6 & 4 & \tabledash & 5 & 4 & \tabledash \\
\ttbar decay $\mu$ & \tabledash & 8.0 & 3.2 & \tabledash & 8.4 & 3.0 \\
\ttW & 10.4 & 2.6 & 18 & 5 & 10.4 & 2.6 & 19 & 4 \\
\ttZ & 5.2 & 1.1 & 5.6 & 1.1 & 5.0 & 1.1 & 5.5 & 1.1 \\
\ttH & 7.6 & 1.3 & 11.4 & 1.9 & 7.5 & 1.3 & 11.2 & 1.8 \\
V EW & 2.4 & 0.9 & 1.9 & 1.5 & 2.3 & 0.8 & 1.5 & 1.2 \\
Top other & 5.4 & 2.2 & 11 & 4 & 5.1 & 2.2 & 11 & 4 \\
\midrule
Total prediction & 52 & 6 & 56 & 4 & 52 & 6 & 56 & 4 \\
\midrule
Data & \multicolumn{1}{l}{\ \ 52} & & \multicolumn{1}{l}{\ \ 51} & & \multicolumn{1}{l}{\ \ 52} & & \multicolumn{1}{l}{\ \ 51} \\
\bottomrule
\end{tabular}
\label{tab:background_yields_unblind}
\end{table}

The upper limits on the signal cross-sections are translated into upper limits on the mixing parameters, using the following equation:
\begin{equation}
\sigma_{\ell,N} = 2 \cdot \sigma_{\ttbar} \cdot |{V_{\ell,N}}|^2 \cdot{\cal B}(W^+\rightarrow N_{|V_{\ell,N}|=1} \ell^+) \cdot{\cal B}(N\rightarrow \ell^+W^-),
\end{equation}

where ${\cal B}(W^+\rightarrow N_{|V_{\ell,N}|=1} \ell^+)$ is the branching ratio of the $W^+ \rightarrow N \ell^+$ decay, for $|V_{\ell,N}|=1$.
The value of ${\cal B}(N\rightarrow \ell^+W^-)$ does not depend on the mixing parameter $V_{\ell,N}$, and varies from 36\% to 39\% in the mass range considered.

\begin{figure}[htbp]
\centering
\subfloat[]{\includegraphics[width=0.48\textwidth]{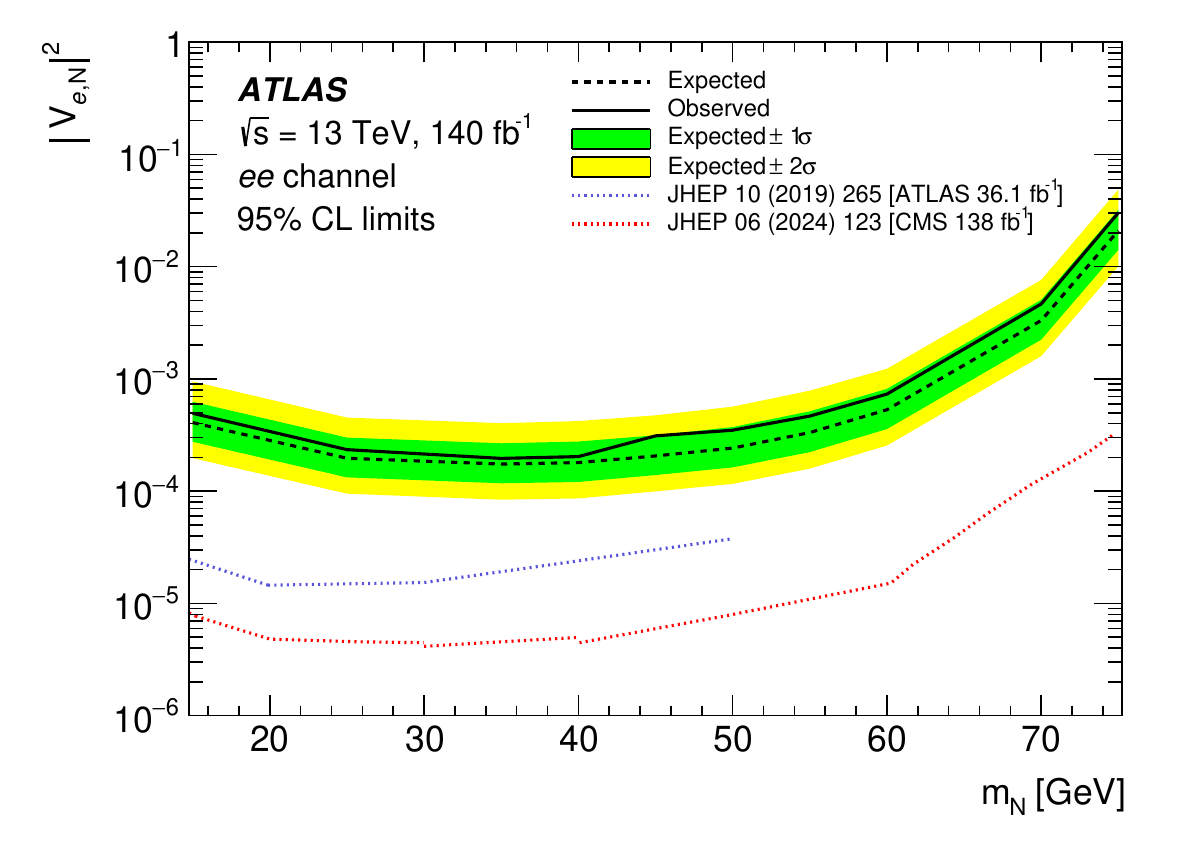}}
\subfloat[]{\includegraphics[width=0.48\textwidth]{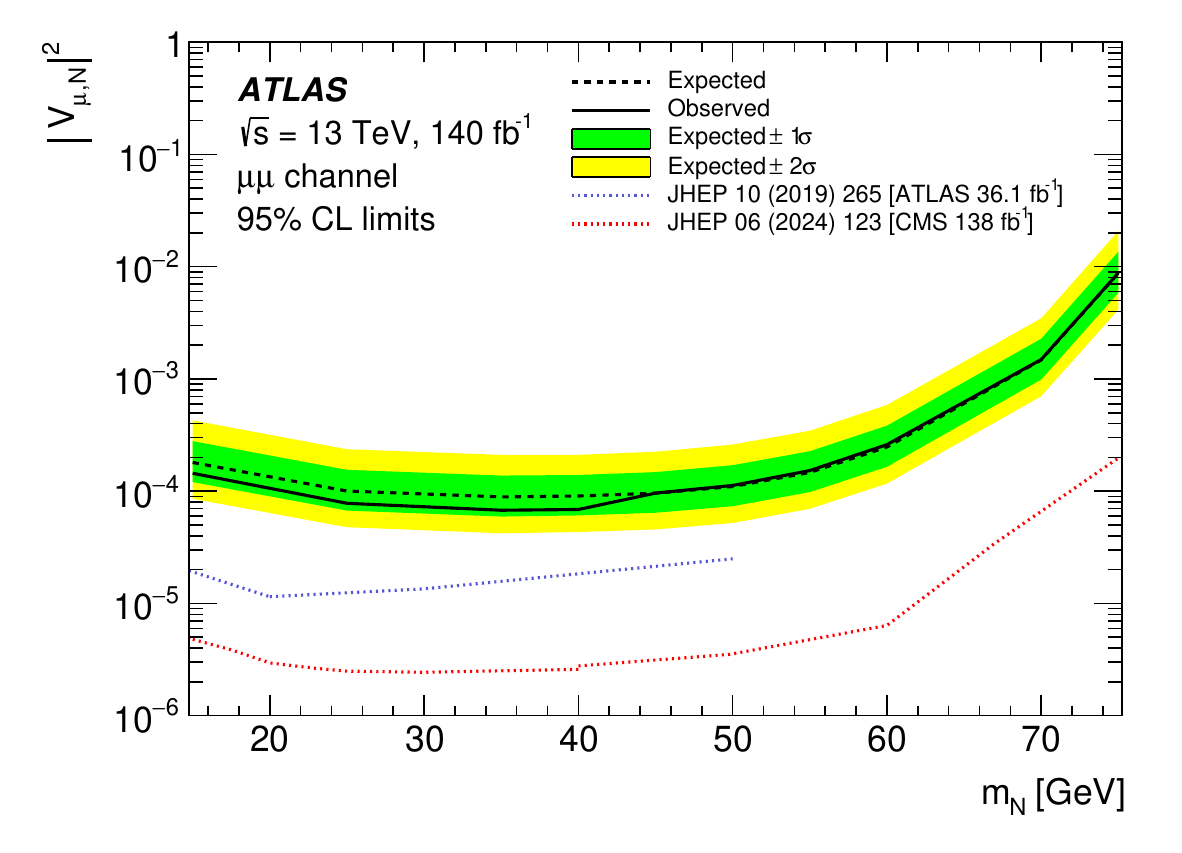}}
\\
\subfloat[]{\includegraphics[width=0.48\textwidth]{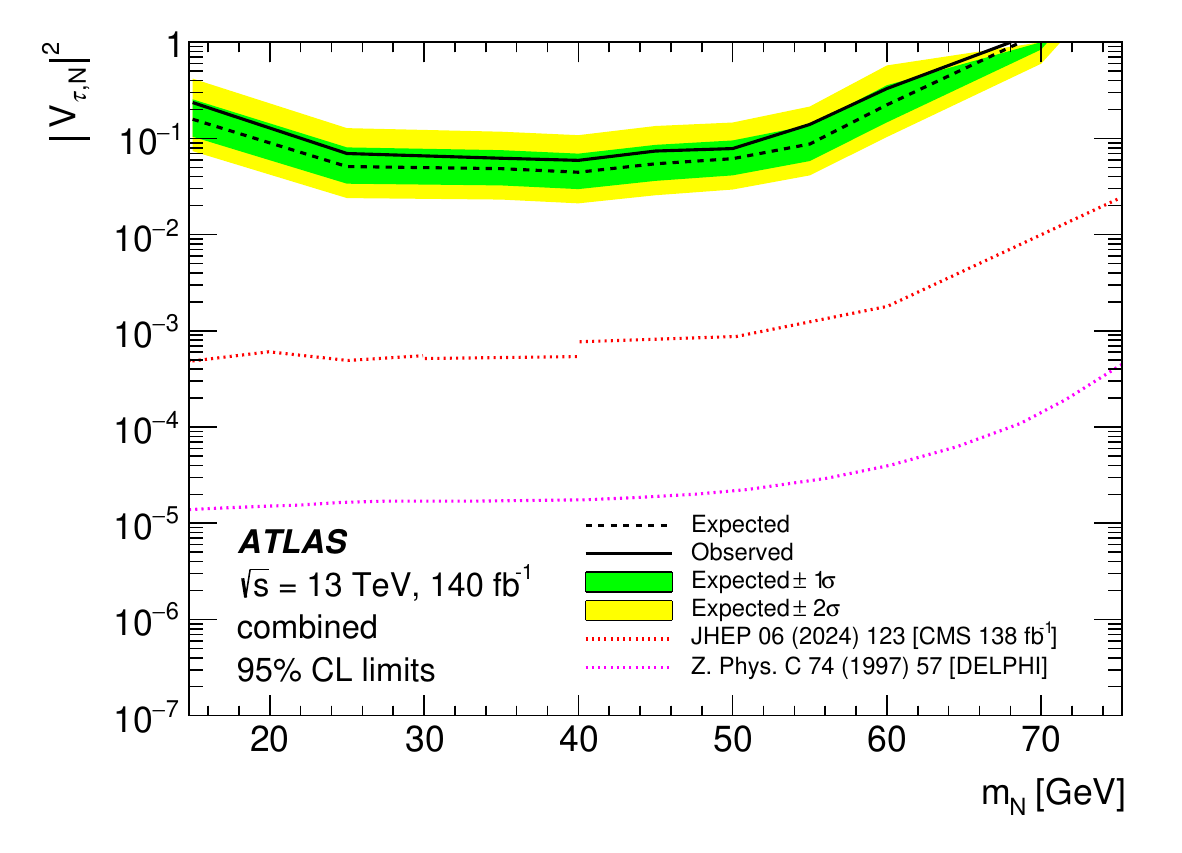}}
\caption{Expected (dashed black lines) and observed (solid black lines) upper limits on the strength of HNL mixing with (a) electron neutrinos, (b) muon neutrinos and (c) $\tau$-neutrinos at 95\% CL, along with the $1\sigma$ (green) and $2\sigma$ (yellow) uncertainty bands. The small discontinuity after 40~\GeV\ reflects the use of a different BDT beyond this mass point. The observed upper limits from some of the previous searches for HNLs in \Wboson boson decays are shown as dotted lines~\cite{DELPHI:1996qcc,CMS-EXO-22-011}.
The discontinuities seen in the result from the CMS Collaboration originate from the different optimization strategies applied in different mass regions.
}
\label{fig:limit_on_coupling_unblind}
\end{figure}

\begin{table}[htbp]
\centering
\caption{Expected and observed upper limits on the signal cross-sections at 95\% CL.}
\begin{tabular}{c S S S S S S S S S S}
\toprule
$m_N$ [$\GeV$] & 15 & 25 & 35 & 40 & 45 & 50 & 55 & 60 & 70 & 75 \\
\midrule
Exp. $\sigma_{e,N}$ [fb] & 21 & 9.8 & 7.3 & 6.9 & 6.9 & 6.7 & 7.2 & 8.5 & 18 & 36 \\
Obs. $\sigma_{e,N}$ [fb] & 26 & 12 & 8.2 & 7.8 & 10 & 9.7 & 10 & 12 & 26 & 52 \\
\midrule
Exp. $\sigma_{\mu,N}$ [fb] & 9.3 & 5.0 & 3.7 & 3.5 & 3.2 & 3.1 & 3.2 & 4.0 & 8.2 & 15 \\
Obs. $\sigma_{\mu,N}$ [fb] & 7.5 & 3.9 & 2.8 & 2.6 & 3.2 & 3.1 & 3.3 & 4.2 & 8.3 & 15 \\
\midrule
Exp. $\sigma_{\tau,N}$ [pb] & 8.9 & 2.6 & 2.1 & 1.7 & 1.8 & 1.8 & 2.0 & 3.7 & 7.0 & 19 \\
Obs. $\sigma_{\tau,N}$ [pb] & 13 & 3.6 & 2.7 & 2.3 & 2.5 & 2.2 & 3.2 & 5.5 & 7.3 & 20 \\
\bottomrule
\end{tabular}
\label{tab:limit_unblind}
\end{table}

\cref{fig:limit_on_coupling_unblind} shows the upper limits on the three mixing parameters as functions of the HNL mass. The observed upper limits on $|V_{e,N}|^2$ and $|V_{\mu,N}|^2$ are comparable across the whole mass range, ${|V_{\mu,N}|}^2$ being slightly more stringent. Nevertheless, these limits are weaker compared with some of the previous results that are shown in this figure. The limits on $|V_{\tau,N}|^2$ are three orders of magnitude weaker than the other two, and cross unity for the mass points above $68~\GeV$. This large difference in sensitivity is primarily because the analysis targets only $ee$ and $\mu\mu$ final states, consequently all the signal events where the same-charge $\tau$-leptons decay into two different flavors of lighter leptons or decay hadronically are not selected by the selection criteria of the SRs.

This study provides complementary information for combinations with other HNL measurements, and may be re-interpreted within the framework of effective field theories (EFT). Dimension-7 (and above) EFT operators, which enable lepton-number violation and complex five-fermion interactions, could produce final states similar to those considered~\cite{PhysRevD.90.125023}.
%

%
%
%
%
%
%
%
%
%
%
%
%
%
%
%
%
%
%
%
%
%
%
%
%
%
%
%
%
%

%
%
%


%
%
\FloatBarrier

\section{Conclusion}
\label{sec:conclusions}
%

A search for heavy neutral leptons produced in \ttbar events with dileptonic final states is performed using 140~\ifb\ of proton--proton collision data collected by the ATLAS detector at the LHC at a center-of-mass energy of 13~\TeV. This is the first search for HNLs using \ttbar events, and uses the distinct signature of same-charge leptons in the final states.

Observations are consistent with the Standard Model predictions, and the search results are presented as 95\% CL upper limits on the mixing parameters $|V_{e,N}|^2$, $|V_{\mu,N}|^2$ and $|V_{\tau,N}|^2$ as functions of the mass of the heavy neutral lepton candidates.

The analysis is optimized to the search for mixing with the first two generations of leptons, and is also used to search for the mixing with the third generation. The best observed upper limits on the mixing parameters in the search region are ${|V_{e,N}|}^2 < 2.0 \times 10^{-4}$, ${|V_{\mu,N}|}^2 < 6.8 \times 10^{-5}$ and ${|V_{\tau,N}|}^2 < 5.9 \times 10^{-2}$. While these upper limits are about one order of magnitude weaker than those obtained in the previous ATLAS search for HNLs in the decays of $W$ bosons~\cite{ATLAS:jhep10.2019.265}, this analysis extends the search region for the first two generations beyond $m_N=50~\GeV$, probing HNL masses up to 75~\GeV, and broadens the scope of previous ATLAS searches by looking for an HNL mixing with the third generation.

Although the sensitivity of this channel to HNL mixing is about 2--3 orders of magnitude lower than channels with \Wboson bosons produced directly in proton--proton collisions, searches for anomalous couplings in top-quark interactions remain a valuable endeavor, as the final states probed here can also be predicted by other theories such as EFT models allowing lepton-number violation and complex five-fermion interactions.


%
%
%
%
%
\section*{Acknowledgements}
%

%

%
%

%
%

We thank CERN for the very successful operation of the LHC and its injectors, as well as the support staff at
CERN and at our institutions worldwide without whom ATLAS could not be operated efficiently.

The crucial computing support from all WLCG partners is acknowledged gratefully, in particular from CERN, the ATLAS Tier-1 facilities at TRIUMF/SFU (Canada), NDGF (Denmark, Norway, Sweden), CC-IN2P3 (France), KIT/GridKA (Germany), INFN-CNAF (Italy), NL-T1 (Netherlands), PIC (Spain), RAL (UK) and BNL (USA), the Tier-2 facilities worldwide and large non-WLCG resource providers. Major contributors of computing resources are listed in Ref.~\cite{ATL-SOFT-PUB-2023-001}.

We gratefully acknowledge the support of ANPCyT, Argentina; YerPhI, Armenia; ARC, Australia; BMWFW and FWF, Austria; ANAS, Azerbaijan; CNPq and FAPESP, Brazil; NSERC, NRC and CFI, Canada; CERN; ANID, Chile; CAS, MOST and NSFC, China; Minciencias, Colombia; MEYS CR, Czech Republic; DNRF and DNSRC, Denmark; IN2P3-CNRS and CEA-DRF/IRFU, France; SRNSFG, Georgia; BMBF, HGF and MPG, Germany; GSRI, Greece; RGC and Hong Kong SAR, China; ISF and Benoziyo Center, Israel; INFN, Italy; MEXT and JSPS, Japan; CNRST, Morocco; NWO, Netherlands; RCN, Norway; MNiSW, Poland; FCT, Portugal; MNE/IFA, Romania; MESTD, Serbia; MSSR, Slovakia; ARIS and MVZI, Slovenia; DSI/NRF, South Africa; MICIU/AEI, Spain; SRC and Wallenberg Foundation, Sweden; SERI, SNSF and Cantons of Bern and Geneva, Switzerland; NSTC, Taipei; TENMAK, T\"urkiye; STFC/UKRI, United Kingdom; DOE and NSF, United States of America.

Individual groups and members have received support from BCKDF, CANARIE, CRC and DRAC, Canada; CERN-CZ, FORTE and PRIMUS, Czech Republic; COST, ERC, ERDF, Horizon 2020, ICSC-NextGenerationEU and Marie Sk{\l}odowska-Curie Actions, European Union; Investissements d'Avenir Labex, Investissements d'Avenir Idex and ANR, France; DFG and AvH Foundation, Germany; Herakleitos, Thales and Aristeia programmes co-financed by EU-ESF and the Greek NSRF, Greece; BSF-NSF and MINERVA, Israel; Norwegian Financial Mechanism 2014-2021, Norway; NCN and NAWA, Poland; La Caixa Banking Foundation, CERCA Programme Generalitat de Catalunya and PROMETEO and GenT Programmes Generalitat Valenciana, Spain; G\"{o}ran Gustafssons Stiftelse, Sweden; The Royal Society and Leverhulme Trust, United Kingdom.

In addition, individual members wish to acknowledge support from CERN: European Organization for Nuclear Research (CERN PJAS); Chile: Agencia Nacional de Investigaci\'on y Desarrollo (FONDECYT 1190886, FONDECYT 1230812, FONDECYT 1230987); China: Chinese Ministry of Science and Technology (MOST-2023YFA1605700), National Natural Science Foundation of China (NSFC - 12175119, NSFC 12275265, NSFC-12075060); Czech Republic: Czech Science Foundation (GACR - 24-11373S), Ministry of Education Youth and Sports (FORTE CZ.02.01.01/00/22\_008/0004632), PRIMUS Research Programme (PRIMUS/21/SCI/017); EU: H2020 European Research Council (ERC - 101002463); European Union: European Research Council (ERC - 948254, ERC 101089007), Horizon 2020 Framework Programme (MUCCA - CHIST-ERA-19-XAI-00), European Union, Future Artificial Intelligence Research (FAIR-NextGenerationEU PE00000013), Italian Center for High Performance Computing, Big Data and Quantum Computing (ICSC, NextGenerationEU); France: Agence Nationale de la Recherche (ANR-20-CE31-0013, ANR-21-CE31-0013, ANR-21-CE31-0022, ANR-22-EDIR-0002), Investissements d'Avenir Labex (ANR-11-LABX-0012); Germany: Baden-Württemberg Stiftung (BW Stiftung-Postdoc Eliteprogramme), Deutsche Forschungsgemeinschaft (DFG - 469666862, DFG - CR 312/5-2); Italy: Istituto Nazionale di Fisica Nucleare (ICSC, NextGenerationEU), Ministero dell'Università e della Ricerca (PRIN - 20223N7F8K - PNRR M4.C2.1.1); Japan: Japan Society for the Promotion of Science (JSPS KAKENHI JP21H05085, JSPS KAKENHI JP22H01227, JSPS KAKENHI JP22H04944, JSPS KAKENHI JP22KK0227); Netherlands: Netherlands Organisation for Scientific Research (NWO Veni 2020 - VI.Veni.202.179); Norway: Research Council of Norway (RCN-314472); Poland: Ministry of Science and Higher Education (IDUB AGH, POB8, D4 no 9722), Polish National Agency for Academic Exchange (PPN/PPO/2020/1/00002/U/00001), Polish National Science Centre (NCN 2021/42/E/ST2/00350, NCN OPUS nr 2022/47/B/ST2/03059, NCN UMO-2019/34/E/ST2/00393, UMO-2020/37/B/ST2/01043, UMO-2021/40/C/ST2/00187, UMO-2022/47/O/ST2/00148, UMO-2023/49/B/ST2/04085); Slovenia: Slovenian Research Agency (ARIS grant J1-3010); Spain: Generalitat Valenciana (Artemisa, FEDER, IDIFEDER/2018/048), Ministry of Science and Innovation (MCIN \& NextGenEU PCI2022-135018-2, MICIN \& FEDER PID2021-125273NB, RYC2019-028510-I, RYC2020-030254-I, RYC2021-031273-I, RYC2022-038164-I), PROMETEO and GenT Programmes Generalitat Valenciana (CIDEGENT/2019/023, CIDEGENT/2019/027); Sweden: Swedish Research Council (Swedish Research Council 2023-04654, VR 2018-00482, VR 2022-03845, VR 2022-04683, VR 2023-03403, VR grant 2021-03651), Knut and Alice Wallenberg Foundation (KAW 2018.0157, KAW 2018.0458, KAW 2019.0447, KAW 2022.0358); Switzerland: Swiss National Science Foundation (SNSF - PCEFP2\_194658); United Kingdom: Leverhulme Trust (Leverhulme Trust RPG-2020-004), Royal Society (NIF-R1-231091); United States of America: U.S. Department of Energy (ECA DE-AC02-76SF00515), Neubauer Family Foundation.

%
%


%
\clearpage
\printbibliography
\clearpage
 
\begin{flushleft}
\hypersetup{urlcolor=black}
{\Large The ATLAS Collaboration}

\bigskip

\AtlasOrcid[0000-0002-6665-4934]{G.~Aad}$^\textrm{\scriptsize 104}$,
\AtlasOrcid[0000-0001-7616-1554]{E.~Aakvaag}$^\textrm{\scriptsize 17}$,
\AtlasOrcid[0000-0002-5888-2734]{B.~Abbott}$^\textrm{\scriptsize 123}$,
\AtlasOrcid[0000-0002-0287-5869]{S.~Abdelhameed}$^\textrm{\scriptsize 119a}$,
\AtlasOrcid[0000-0002-1002-1652]{K.~Abeling}$^\textrm{\scriptsize 56}$,
\AtlasOrcid[0000-0001-5763-2760]{N.J.~Abicht}$^\textrm{\scriptsize 50}$,
\AtlasOrcid[0000-0002-8496-9294]{S.H.~Abidi}$^\textrm{\scriptsize 30}$,
\AtlasOrcid[0009-0003-6578-220X]{M.~Aboelela}$^\textrm{\scriptsize 45}$,
\AtlasOrcid[0000-0002-9987-2292]{A.~Aboulhorma}$^\textrm{\scriptsize 36e}$,
\AtlasOrcid[0000-0001-5329-6640]{H.~Abramowicz}$^\textrm{\scriptsize 154}$
% ...
\bigskip
\\

% ...
$^{17}$Department for Physics and Technology, University of Bergen, Bergen; Norway.\\
% ...
$^{30}$Physics Department, Brookhaven National Laboratory, Upton NY; United States of America.\\
% ...
$^{36}$$^{(a)}$Facult\'e des Sciences Ain Chock, Universit\'e Hassan II de Casablanca;$^{(b)}$Facult\'{e} des Sciences, Universit\'{e} Ibn-Tofail, K\'{e}nitra;$^{(c)}$Facult\'e des Sciences Semlalia, Universit\'e Cadi Ayyad, LPHEA-Marrakech;$^{(d)}$LPMR, Facult\'e des Sciences, Universit\'e Mohamed Premier, Oujda;$^{(e)}$Facult\'e des sciences, Universit\'e Mohammed V, Rabat;$^{(f)}$Institute of Applied Physics, Mohammed VI Polytechnic University, Ben Guerir; Morocco.\\
% ...
$^{45}$Physics Department, Southern Methodist University, Dallas TX; United States of America.\\
% ...
$^{50}$Fakult\"{a}t Physik , Technische Universit{\"a}t Dortmund, Dortmund; Germany.\\
% ...
$^{56}$II. Physikalisches Institut, Georg-August-Universit\"{a}t G\"ottingen, G\"ottingen; Germany.\\
% ...
$^{104}$CPPM, Aix-Marseille Universit\'e, CNRS/IN2P3, Marseille; France.\\
% ...
$^{119}$$^{(a)}$New York University Abu Dhabi, Abu Dhabi;$^{(b)}$United Arab Emirates University, Al Ain; United Arab Emirates.\\
% ...
$^{123}$Homer L. Dodge Department of Physics and Astronomy, University of Oklahoma, Norman OK; United States of America.\\
% ...
$^{154}$Raymond and Beverly Sackler School of Physics and Astronomy, Tel Aviv University, Tel Aviv; Israel.\\
% ...
$^{a}$ Also Affiliated with an institute covered by a cooperation agreement with CERN.\\
$^{b}$ Also at An-Najah National University, Nablus; Palestine.\\
$^{c}$ Also at Borough of Manhattan Community College, City University of New York, New York NY; United States of America.\\
$^{d}$ Also at Center for Interdisciplinary Research and Innovation (CIRI-AUTH), Thessaloniki; Greece.\\
$^{e}$ Also at Centro Studi e Ricerche Enrico Fermi; Italy.\\
$^{f}$ Also at CERN, Geneva; Switzerland.\\
% ...

\end{flushleft}


%

%
%

%

%
%

\clearpage
\section*{HepData material}
%

\begin{table}[h]
\centering
\small
\caption{The cutflow table for $N1$ and $N2$ signal samples for different mass points. These samples contain both the $ee$ and $\mu\mu$ final states in equal ratio, and each of these samples is normalized to a total cross-section of 0.1~pb.\\
Abbreviations: SC = Same Charge, SF = Same Flavor.}
\resizebox{\textwidth}{!}{%
\begin{tabular}{p{3.5cm} *{10}{r}}
\toprule
\textbf{$m_\mathrm{HNL}$ [GeV]} & \textbf{15} & \textbf{25} & \textbf{35} & \textbf{40} & \textbf{45} & \textbf{50} & \textbf{55} & \textbf{60} & \textbf{70} & \textbf{75} \\
\midrule
\text{$N_\mathrm{initial}$ ($\mathcal{L} \times \sigma$)} & 14000 & 14000 & 14000 & 14000 & 14000 & 14000 & 14000 & 14000 & 14000 & 14000 \\
Trigger selection & 6743 & 6816 & 6592 & 6436 & 6135 & 5876 & 5433 & 4964 & 4252 & 4205 \\
\text{$\ge 2$ $e/\mu$ with $\pt > 10~\GeV$} & 2047 & 2758 & 3059 & 3102 & 3075 & 3006 & 2792 & 2484 & 1669 & 1293 \\
\text{$\ge 3$ jets with $\pt > 20~\GeV$} & 1869 & 2550 & 2847 & 2894 & 2873 & 2815 & 2613 & 2321 & 1545 & 1187 \\
\text{$\ge$ 2 same-charge leptons} & 755 & 1168 & 1356 & 1387 & 1388 & 1366 & 1253 & 1094 & 631 & 408 \\
\text{$\ge$ 2 $b$-jets with $\pt > 25~\GeV$} & 686 & 1051 & 1084 & 863 & 854 & 845 & 776 & 672 & 391 & 250 \\
\text{$\ge$ 4 non-$b$-tagged jets} & 171 & 279 & 448 & 639 & 641 & 641 & 589 &518 & 293 & 186 \\
\text{Exactly 2 SC SF $e$ or $\mu$} & 145 & 233 & 286 & 304 & 306 & 304 & 276 & 238 & 130 & 79 \\
\bottomrule
\end{tabular}%
}
\label{tab:cutflow_N1_N2}
\end{table}

\begin{table}[h]
\centering
\small
\caption{The cutflow table for $N3$ signal samples for different mass points. These samples contain both the $ee$ and $\mu\mu$ final states in equal ratio, and each of these samples is normalized to a total cross-section of 10~pb.\\
Abbreviations: SS = Same Charge, SF = Same Flavor.}
\resizebox{\textwidth}{!}{%
\begin{tabular}{p{3.5cm} *{10}{r}}
\toprule
\textbf{$m_\mathrm{HNL}$ [GeV]} & \textbf{15} & \textbf{25} & \textbf{35} & \textbf{40} & \textbf{45} & \textbf{50} & \textbf{55} & \textbf{60} & \textbf{70} & \textbf{75} \\
\midrule
\text{$N_\mathrm{initial}$ ($\mathcal{L} \times \sigma$)} & 1400000 & 1400000 & 1400000 & 1400000 & 1400000 & 1400000 & 1400000 & 1400000 & 1400000 & 1400000 \\
\text{Trigger selection} & 164160 & 164583 & 167843 & 164509 & 151691 & 151481 & 140810 & 137873 & 137771 & 111429 \\
\text{$\ge 2$ $e/\mu$ with $\pt > 10~\GeV$} & 6502 & 6898 & 6960 & 6692 & 5990 & 5624 & 4757 & 4113 & 3018 & 2226 \\
\text{$\ge 3$ jets with $\pt > 20~\GeV$} & 5769 & 6172 & 6300 & 6091 & 5498 & 5173 & 4408 & 3811 & 2793 & 2068 \\
\text{$\ge$ 2 same-charge leptons} & 430 & 562 & 716 & 730 & 712 & 677 & 573 & 481 & 218 & 141 \\
\text{$\ge$ 2 $b$-jets with $\pt > 25~\GeV$} & 384 & 508 & 634 & 646 & 634 & 604 & 518 & 434 & 188 & 126 \\
\text{$\ge$ 4 non-$b$-tagged jets} & 76 & 120 & 179 & 187 & 166 & 179 & 153 & 131 & 59 & 40 \\
\text{Exactly 2 SC SF $e$ or $\mu$} & 63 & 103 & 138 & 152 & 130 & 138 & 124 & 94 & 48 & 32 \\
\bottomrule
\end{tabular}%
}
\label{tab:cutflow_N3}
\end{table}


\clearpage

\end{document}